# Compact Finite Difference Scheme with Hermite Interpolation for Pricing American Put Options Based on Regime Switching Model

**Chinonso I. Nwankwo[a], Weizhong Dai[a,*], Ruihua Liu[b]**

[a] Department of Mathematics and Statistics, Louisiana Tech University, Ruston LA 71272, USA
[b] Department of Mathematics, University of Dayton, 300 College Park, Dayton, OH 45469, USA
* Corresponding author, dai@coes.latech.edu

**Abstract**

We consider a system of coupled free boundary problems for pricing American put options with regime-switching. To solve this system, we first employ the logarithmic transformation to map the free boundary for each regime to multi-fixed intervals and then eliminate the first-order derivative in the transformed model by taking derivatives to obtain a system of partial differential equations which we call the asset-delta-gamma-speed equations. As such, the fourth-order compact finite difference scheme can be used for solving this system. The influence of other asset, delta, gamma, and speed options in the present regime is estimated based on Hermite interpolations. Finally, the numerical method is tested with several examples. Our results show that the scheme provides an accurate solution that is fast in computation as compared with other existing numerical methods.

## 1. Introduction

The well-known Black-Scholes model has been used over decades in options valuation. This model constructs a delta hedging portfolio with an assumption of the frictionless market, no-arbitrage, and constant risk-free interest and volatility (Ugur, 2006). To remove this ideal assumption and reproduce the actual market price, risk, behavior, and dynamics, researchers have proposed several improvements by including stochastic volatility (Chockalingam and Muthuraman, 2011; Düring and Fournié, 2012; Garnier and Sølna, 2017; Huang et al., 2011; Hull and White, 1987; Ikonen and Toivanen, 2007; Zhylyevskyy, 2009), jump-diffusion (Bingham, 2006; Chen et al., 2019; Cont and Tankov, 2004; Guoqing and Hanson, 2006; Kou, 2002), and regime-switching (Company et al., 2016a; Egorova et al., 2016; Huang et al., 2011; Khaliq and Liu, 2009; Mamon and Rodrigo, 2005) in the pricing models.

The regime-switching model for American option valuation, first introduced by Hamilton (1989), has gained broader interest after the seminal work of Buffington and Elliot (2002). It defines a finite number of market states known as regimes. Each regime has its own set of market variables, and the market randomly switches among different regimes (Chiarella et al., 2016). The model for option valuation with regime-switching involves a system of partial differential equations with free boundaries for which the analytical solution is very difficult to obtain in general. Thus, some works in the literature have proposed numerical techniques for solving the option pricing equation with regime-switching. Among them, the commonly known numerical methods are the penalty method (Khaliq and Liu, 2009; Nielsen et al., 2002; Zhang et al., 2013), the method of line (MOL) (Chiarella et al., 2016; Meyer and van der Hoek, 1997), the lattice method (Han and Kim, 2016; Shang and Bryne, 2019), the fast Fourier transform (Boyarchenko and Levendorskii, 2008; Liu et al., 2006), and the front-fixing techniques (Egorova et al., 2016). The lattice-based method is more common among practitioners. However, tracking the optimal exercise boundary can be a challenge (Shang and Bryne, 2019). Fast Fourier transform method is efficient in solving the European options (Chiarella et al., 2016). The penalty method removes the free boundary by introducing a penalty term (Khaliq and Liu, 2009). The MOL method calculates the asset and delta options and the optimal exercise boundary simultaneously during computation. Meyer and van der Hoek (1997) pointed out that there are still some complications with the MOL method due to the singularity of the solution and infinite interval. The front-fixing technique (Blackwell and Hogan, 1994; Company et al., 2016b; Company et al., 2016c; Landau, 1950; Mitchell and Vynnycky, 2009; Mitchell and Vynnycky) was first applied by Egorova et al. (2016) to the regime-switching model.

To the best of our knowledge, the above methods provide up to second-order accurate solutions. The motivation of this research is to propose a higher accurate front-fixing numerical method for solving the regime-switching pricing model. To this end, we first use a logarithmic transformation to map the free boundary for each regime to multi-fixed intervals and then eliminate the first-order derivative in the transformed model by taking derivatives to obtain a system of partial differential equations which we call the asset-delta-gamma-speed equations. As such, the fourth-order compact finite difference can be used for solving this system. The influence of other asset, delta, gamma, speed options in the present regime is estimated based on Hermite interpolations. Finally, the numerical scheme is solved using either the Gauss-Seidel or Newton iterative method, which predicts the optimal exercise boundary, option value, and option Greeks in each regime.

The rest of the paper is organized as follows. In section 2, we consider a regime-switching model and its transformations. We transform the model to obtain coupled partial differential equations for option values, delta, gamma, and speed options in each regime. In section 3, we develop an accurate numerical method and its algorithms for solving these equations and obtaining the option values, optimal exercise boundary, and the Greeks in each regime. In section 4, we test our algorithms using examples with two, four, eight, and sixteen regimes. We conclude the paper in section 5.

## 2. Regime Switching Model and its Transformations

### 2.1. Regime Switching Model

Let us consider a continuous-time Markov chain whose states are labeled as $m = 1,2,\cdots,I$. Let $Q = (q_{ml})_{I \times I}$ represent the generator matrix with the entry elements $q_{ml}$ satisfying the condition below (Norris, 1998):

$$q_{mm} = -\sum_{l \neq m} q_{ml}, \quad q_{ml} \geq 0, \quad for\ l \neq m, \quad l = 1,2,\cdots,I. \tag{1}$$

Assuming a risk-neutral measure (Elliot et al., 2007), the underlying asset follows a stochastic process

$$dS_t = S_t(r_{\alpha_t} dt + \sigma_{\alpha_t} dB_t), \quad 0 \leq t < \infty, \tag{2}$$

where $r_{\alpha_t}$ and $\sigma_{\alpha_t}$ are the interest rate and volatility of the asset, respectively, and are dependent on the Markov chain state with

$$r_{\alpha_t|\alpha_t} = r_m, \quad \sigma_{\alpha_t|\alpha_t} = \sigma_m, \quad m = 1,2,\cdots,I. \tag{3}$$

We consider an American put option written on the asset $S_t$ with strike price $K$ and expiration time $T$. Let $V_m(S,t)$ denote the option price and $\tau = T - t$. Then, $V_m(S,\tau)$ satisfies the following parabolic PDEs with free boundaries (Khaliq and Liu, 2009):

$$-\frac{\partial V_m(S,\tau)}{\partial \tau} + \frac{1}{2}\sigma^2{}_m S^2 \frac{\partial^2 V_m(S,\tau)}{\partial S^2} + r_m S \frac{\partial V_m(S,\tau)}{\partial S} - r_m V_m(S,\tau) + \sum_{l \neq m} q_{ml}[V_l(S,\tau) - V_m(S,\tau)] = 0,$$

$$for\ S > s_{f(m)}(\tau), \tag{4}$$

$$V_m(S,\tau) = K - S, \quad for\ S < s_{f(m)}(\tau). \tag{5}$$

Here, the initial and boundary conditions are given as:

$$V_m(S,0) = max(K - S, 0), \quad s_{f(m)}(0) = K; \tag{6}$$

$$V_m\big(s_{f(m)},\tau\big) = K - s_{f(m)}(\tau), \quad V_m(0,\tau) = K, \quad V_m(\infty,\tau) = 0, \quad \frac{\partial}{\partial S}V_m\big(s_{f(m)},\tau\big) = -1, \tag{7}$$

where $s_{f(m)}(\tau)$ is the optimal exercise boundary for the $m^{th}$ regime.

## 2.2. Logarithmic Transformation

To fix the free boundary challenge, we employ a transformation (Egorova et al., 2016; Wu and Kwok, 1997) on multi-variable domains as

$$x_m = \ln\frac{S}{s_{f(m)}(\tau)} = \ln S - \ln s_{f(m)}(\tau), \qquad m = 1,2,\cdots,I, \tag{8}$$

where the variable $x_m$ exists in a positive domain $x_m \in (0,\infty)$ if $S > s_{f(m)}$. The transformed $m$ option value functions $U_m(x_m,\tau)$ are related to the original $m$ option value functions $V_m(S,\tau)$ by the dimensionless transformation

$$U_m(x_m,\tau) = V_m(S,\tau), \qquad m = 1,2,\cdots,I. \tag{9a}$$

Applying this transformation, we obtain the following relations:

$$\frac{\partial x_m}{\partial S} = \frac{1}{S}, \quad \frac{\partial x_m}{\partial \tau} = -\frac{s'_{f(m)}(\tau)}{s_{f(m)}(\tau)}, \quad \frac{\partial V_m}{\partial S} = \frac{1}{S}\frac{\partial U_m}{\partial x_m}; \tag{9b}$$

$$\frac{\partial^2 V_m}{\partial S^2} = \frac{1}{S^2}\left(-\frac{\partial U_m}{\partial x_m} + \frac{\partial^2 U_m}{\partial x_m^2}\right), \quad \frac{\partial V_m}{\partial \tau} = \frac{\partial U_m}{\partial \tau} - \frac{s'_{f(m)}(\tau)}{s_{f(m)}(\tau)}\frac{\partial U_m}{\partial x_m}. \tag{9c}$$

Because our interest is to also calculate speed, delta decay, and color options, we differentiate further to obtain higher derivatives of the $m$th option value function as

$$\frac{\partial^3 V_m}{\partial S^3} = \frac{1}{S^3}\left(2\frac{\partial U_m}{\partial x_m} - 3\frac{\partial^2 U_m}{\partial x_m^2} + \frac{\partial^3 U_m}{\partial x_m^3}\right), \quad \frac{\partial^2 V_m}{\partial S\partial\tau} = \frac{1}{S}\left(\frac{\partial^2 U_m}{\partial x_m \partial \tau} - \frac{s'_{f(m)}(\tau)}{s_{f(m)}(\tau)}\frac{\partial^2 U_m}{\partial x_m^2}\right); \tag{9d}$$

$$\frac{\partial^3 V_m}{\partial S^2 \partial \tau} = \frac{1}{S^2}\left(\frac{\partial^3 U_m}{\partial x_m^2 \partial \tau} - \frac{\partial^2 U_m}{\partial x_m \partial \tau} + \frac{s'_{f(m)}(\tau)}{s_{f(m)}(\tau)}\frac{\partial^2 U_m}{\partial x_m^2} - \frac{s'_{f(m)}(\tau)}{s_{f(m)}(\tau)}\frac{\partial^3 U_m}{\partial x_m^3}\right). \tag{9e}$$

Let $l$ represent the coupled regime(s) in the $m$ free boundary PDEs. The former also has a variable

$$x_l = \ln\frac{S}{s_{f(l)}(\tau)} = \ln S - \ln s_{f(l)}(\tau), \qquad l \neq m, \qquad l = 1,2,\cdots,I. \tag{10}$$

Eliminating $S$ in the $l^{th}$ and $m^{th}$ equations, we obtain

$$x_l = x_m - \ln\frac{s_{f(l)}(\tau)}{s_{f(m)}(\tau)}. \tag{11}$$

Substituting (8) into (4) and (5) (i.e., $S = s_{f(m)}(\tau)e^{x_m}$), the model can be changed to

$$\frac{\partial U_m(x_m,\tau)}{\partial \tau} - \frac{1}{2}\sigma^2_m \frac{\partial^2 U_m(x_m,\tau)}{\partial x_m^2} - \left(\frac{s'_{f(m)}}{s_{f(m)}} + r_m - \frac{\sigma^2_m}{2}\right)\frac{\partial U_m(x_m,\tau)}{\partial x_m} + r_m U_m(x_m,\tau)$$

$$- \sum_{l \neq m} q_{ml}\left[U_l(x_m,\tau) - U_m(x_m,\tau)\right] = 0, \quad \text{for } x_m > 0; \quad (12)$$

$$U_m(x_m,\tau) = K - S = K - s_{f(m)}(\tau)e^{x_m}, \quad \text{for } x_m < 0; \quad (13)$$

where the initial condition (5) is changed to

$$U_m(x_m,0) = \max(K - Ke^{x_m}, 0) = 0, \quad x_m \geq 0, \quad s_{f(m)}(0) = K. \quad (14)$$

By letting $x_m \to 0^-$, we obtain from (13) that $U_m(0,\tau) = K - s_{f(m)}(\tau)$. Thus, together with (7), we obtain the boundary condition for (12) as

$$U_m(0,\tau) = K - s_{f(m)}(\tau), \quad U_m(\infty,\tau) = 0. \quad (15)$$

To apply the high order compact finite difference method, we further transform the system in (12)-(15) by eliminating the first-order derivative. To this end, we let $W_m(x_m,\tau)$ represent the derivative of the option value in each regime known as the delta option and given as

$$W_m(x_m,\tau) = \frac{\partial U_m(x_m,\tau)}{\partial x_m}, \quad \text{for } x_m > 0 \text{ and } x_m < 0. \quad (16)$$

Differentiating (12) and (13) with respect to $x_m$, respectively, we generate a system of partial differential equations in terms of delta option as

$$\frac{\partial W_m(x_m,\tau)}{\partial \tau} - \frac{1}{2}\sigma^2_m \frac{\partial^2 W_m(x_m,\tau)}{\partial x_m^2} - \left(\frac{s'_{f(m)}}{s_{f(m)}} + r_m - \frac{\sigma^2_m}{2}\right)\frac{\partial^2 U_m(x_m,\tau)}{\partial x_m^2} + r_m W_m(x_m,\tau)$$

$$- \sum_{l \neq m} q_{ml}\left(W_l(x_m,\tau) - W_m(x_m,\tau)\right) = 0, \quad \text{for } x_m > 0; \quad (17)$$

$$W_m(x_m,\tau) = -s_{f(m)}e^{x_m}, \quad \text{for } x_m < 0; \quad (18)$$

where the initial condition for (17) is obtained based on (14) as:

$$W_m(x_m,0) = 0, \quad x_m \geq 0. \quad (19)$$

By letting $x_m \to 0^-$ in (18) together with (7), we obtain the boundary condition for (17) as

$$W_m(0,\tau) = -s_{f(m)}, \quad W_m(\infty,\tau) = 0. \quad (20)$$

It is important to point out that for $W_m(x_m,\tau)$ at $x_m = 0$ when $\tau = 0$, its value obtained based on the initial condition and the boundary condition are different. This happens in many PDE problems. We are mostly concerned with what happens for $W_m(x_m, \tau)$ and other functions in $x_m \geq 0$ when $\tau > 0$. Here, we set $W_m(0,0) = 0$. We have used the average of the limit from the left and right as the value of $W_m(0,0)$. We have also used the smoothstep method (Bravo and Mcgraw, 2015) to approximate $W_m(0,0)$. However, when comparing them with our choice of $W_m(0,0) = 0$, we found no significant difference.

Furthermore, we let $Y_m(x_m, \tau)$ represent the derivative of the delta option in the $m^{th}$ regime known as the gamma option and obtain

$$Y_m(x_m, \tau) = \frac{\partial W_m(x_m, \tau)}{\partial x_m} = \frac{\partial^2 U_m(x_m, \tau)}{\partial x_m^2}, \quad \text{for } x_m > 0 \text{ and } x_m < 0. \tag{21}$$

Differentiating (17) and (18) with respect to $x_m$, respectively, we generate a system of gamma option PDEs for each regime of the form

$$\frac{\partial Y_m(x_m, \tau)}{\partial \tau} - \frac{1}{2}\sigma^2_m \frac{\partial^2 Y_m(x_m, \tau)}{\partial x_m^2} - \left(\frac{s'_{f(m)}}{s_{f(m)}} + r_m - \frac{\sigma^2_m}{2}\right)\frac{\partial^2 W_m(x_m, \tau)}{\partial x_m^2} + r_m Y_m(x_m, \tau)$$

$$- \sum_{l \neq m} q_{ml}\left(Y_l(x_m, \tau) - Y_m(x_m, \tau)\right) = 0, \quad \text{for } x_m > 0; \tag{22}$$

$$Y_m(x_m, \tau) = -s_{f(m)} e^{x_m}, \quad \text{for } x_m < 0; \tag{23}$$

where the initial condition for (22) is obtained based on (19) as

$$Y_m(x_m, 0) = 0, \quad x_m \geq 0. \tag{24}$$

By letting $x_m \to 0^-$ in (23) together with (7), we obtain the boundary condition for (22)

$$Y_m(0, \tau) = -s_{f(m)}, \quad Y_m(\infty, \tau) = 0. \tag{25}$$

Finally, we let $Z_m(x_m, \tau)$ represent the derivative of the gamma option known as the speed option and obtain

$$Z_m(x_m, \tau) = \frac{\partial Y_m(x_m, \tau)}{\partial x_m} = \frac{\partial^2 W_m(x_m, \tau)}{\partial x_m^2} = \frac{\partial^2 U_m(x_m, \tau)}{\partial x_m^2}, \quad \text{for } x_m > 0 \text{ and } x_m < 0, \tag{26}$$

Differentiating (22), (23) with respect to $x_m$, we generate a system of speed option PDEs as

$$\frac{\partial Z_m(x_m, \tau)}{\partial \tau} - \frac{1}{2}\sigma^2_m \frac{\partial^2 Z_m(x_m, \tau)}{\partial x_m^2} - \left(\frac{s'_{f(m)}}{s_{f(m)}} + r_m - \frac{\sigma^2_m}{2}\right)\frac{\partial^2 Y_m(x_m, \tau)}{\partial x_m^2} + r_m Z_m(x_m, \tau)$$

$$- \sum_{l \neq m} q_{ml}\left(Z_l(x_m, \tau) - Z_m(x_m, \tau)\right) = 0, \quad \text{for } x_m > 0; \tag{27}$$

$$Z_m(x_m, \tau) = -s_{f(m)} e^{x_m}, \quad \text{for } x_m < 0; \tag{28}$$

where the initial and boundary conditions for (27) are given as:

$$Z_m(x_m, 0) = 0, \quad x_m \geq 0, \quad Z_m(0, \tau) = -s_{f(m)}, \quad Z_m(\infty, \tau) = 0. \tag{29}$$

Thus, a set of asset-delta-gamma-speed option PDEs in each regime can be written as follows:

$$\frac{\partial U_m}{\partial \tau} - \frac{1}{2}\sigma^2_m \frac{\partial^2 U_m}{\partial x_m^2} - \left(\frac{s'_{f(m)}}{s_{f(m)}} + r_m - \frac{\sigma^2_m}{2}\right) W_m + (r_m - q_{mm})U_m - \sum_{l \neq m} q_{ml} U_l = 0, \tag{30a}$$

$$\frac{\partial W_m}{\partial \tau} - \frac{1}{2}\sigma^2_m \frac{\partial^2 W_m}{\partial x_m^2} - \left(\frac{s'_{f(m)}}{s_{f(m)}} + r_m - \frac{\sigma^2_m}{2}\right)\frac{\partial^2 U_m}{\partial x_m^2} + (r_m - q_{mm})W_m - \sum_{l \neq m} q_{ml} W_l = 0, \tag{30b}$$

$$\frac{\partial Y_m}{\partial \tau} - \frac{1}{2}\sigma^2_m \frac{\partial^2 Y_m}{\partial x_m^2} - \left(\frac{s'_{f(m)}}{s_{f(m)}} + r_m - \frac{\sigma^2_m}{2}\right)\frac{\partial^2 W_m}{\partial x_m^2} + (r_m - q_{mm})Y_m - \sum_{l \neq m} q_{ml} Y_l = 0, \tag{30c}$$

$$\frac{\partial Z_m}{\partial \tau} - \frac{1}{2}\sigma^2_m \frac{\partial^2 Z_m}{\partial x_m^2} - \left(\frac{s'_{f(m)}}{s_{f(m)}} + r_m - \frac{\sigma^2_m}{2}\right)\frac{\partial^2 Y_m}{\partial x_m^2} + (r_m - q_{mm})Z_m - \sum_{l \neq m} q_{ml} Z_l = 0, \tag{30d}$$

where $m = 1, 2, \cdots, I$, $x_m > 0$, and the initial and boundary conditions for $U_m(x_m, \tau)$, $W_m(x_m, \tau)$, $Y_m(x_m, \tau)$, and $Z_m(x_m, \tau)$ are given as:

$$U_m(x_m, 0) = 0, \quad W_m(x_m, 0) = 0, \quad Y_m(x_m, 0) = 0; \tag{31a}$$

$$Z_m(x_m, 0) = 0, \quad s_{f(m)}(0) = K, \quad x_m \geq 0; \tag{31b}$$

$$U_m(0, \tau) = K - s_{f(m)}(\tau), \quad W_m(0, \tau) = Y_m(0, \tau) = Z_m(0, \tau) = -s_{f(m)}(\tau); \tag{31c}$$

$$U_m(\infty, \tau) = 0; \quad W_m(\infty, \tau) = 0, \quad Y_m(\infty, \tau) = 0, \quad Z_m(\infty, \tau) = 0. \tag{31d}$$

On the other hand, for $x_m < 0$,

$$U_m(x_m, \tau) = K - s_{f(m)}(\tau)e^{x_m}, \quad W_m(x_m, \tau) = -s_{f(m)}(\tau)e^{x_m}; \tag{32a}$$

$$Y_m(x_m, \tau) = -s_{f(m)}(\tau)e^{x_m}, \quad Z_m(x_m, \tau) = -s_{f(m)}(\tau)e^{x_m}. \tag{32b}$$

## 3. Numerical Formulation

To solve the above asset-delta-gamma-speed option PDEs, we first design a uniform grid $[0, \infty) \times [0\ T]$ for each regime taking into consideration how the $m^{th}$ regime's interval relates to the $l^{th}$ regime's interval using the Hermite interpolation technique. The infinite boundary is replaced with the far estimate boundary (Egorova et al., 2016; Kangro and Nicolaides, 2000; Toivanen, 2010), which we denote as $(x_m)_M$. Representing $i$ as the node point in the $m^{th}$ regime's interval, $j$ as the node point in the $l^{th}$ regime's interval and $n$ as the time level. For given positive integers M and N representing the numbers of grid points and time steps, respectively, we have

$$(x_m)_i = ih, \quad (x_l)_j = jh, \quad \tau_n = nk, \quad h = \frac{(x_m)_M}{M}, \quad k = \frac{T}{N}, \quad i,j \in [0, M], \quad k \in [0, N]. \tag{33}$$

We denote the numerical solutions of $U_m((x_m)_i, \tau_n), U_l((x_l)_j, \tau_n), W_m((x_m)_i, \tau_n), W_l((x_l)_j, \tau_n)$, $Y_m((x_m)_i, \tau_n), Y_l((x_l)_j, \tau_n), Z_m((x_m)_i, \tau_n), Z_l((x_l)_j, \tau_n), s_{f(m)}(\tau_n),$ and $s_{f(l)}(\tau_n)$ as $(u_m)_i^n, (u_l)_j^n$, $(w_m)_i^n, (w_l)_j^n, (y_m)_i^n, (y_l)_j^n, (z_m)_i^n, (z_l)_j^n, s_{f(m)}^n,$ and $s_{f(l)}^n$, respectively.

### 3.1. Compact Finite Difference Scheme

In the numerical discretization for the asset, delta, gamma and speed options in each regime, the higher-order compact finite difference method is used in space, while the second-order Crank-Nicolson method is used in time. To develop a compact finite difference scheme in space at $(x_m)_0 = 0$, we first derive a compact finite difference formula as described in the following lemma.

**Lemma**. Assume $f(x) \in C^6[x_0, x_1]$, then it holds

$$\frac{7}{4}f''(x_0) + \frac{3}{4}f''(x_1) = \frac{5}{h^2}[f(x_1) - f(x_0)] - \frac{5}{h}f'(x_0) - \frac{h}{4}f^{(3)}(x_0) + \frac{h}{6}f^{(3)}(x_1) + O(h^4). \tag{34}$$

**Proof**. Applying the Taylor expansion at $x_0$, we obtain

$$\frac{h}{12}f''(x_1) - \frac{h}{12}f''(x_0) = \frac{h^2}{12}f^{(3)}(x_0) + \frac{h^3}{24}f^{(4)}(x_0) + \frac{h^4}{72}f^{(5)}(x_0) + \cdots, \tag{35a}$$

$$\frac{5}{3}\left[\frac{f(x_1) - f(x_0)}{h}\right] - \frac{5}{3}f'(x_0) - \frac{5h}{6}f''(x_0) = \frac{5h^2}{18}f^{(3)}(x_0) + \frac{5h^3}{72}f^{(4)}(x_0) + \frac{h^4}{72}f^{(5)}(x_0) + \cdots. \tag{35b}$$

Eliminating the fifth-order derivative by subtracting (35b) from (35a), we obtain

$$-\frac{5}{3}\left[\frac{f(x_1) - f(x_0)}{h}\right] + \frac{5}{3}f'(x_0) + \frac{h}{12}f''(x_1) + \frac{9h}{12}f''(x_0) = -\frac{7h^2}{36}f^{(3)}(x_0) - \frac{h^3}{36}f^{(4)}(x_0) + O(h^5). \tag{36}$$

On the other hand, we have (Hirsh, 1975)

$$\frac{2}{3}f'(x_0) + \frac{1}{3}f'(x_1) = \left[\frac{f(x_1) - f(x_0)}{h}\right] - \frac{h}{6}f''(x_0) + O(h^3), \tag{37a}$$

for which we multiply it by $h^2/6$, differentiate twice, and then rearrange. This gives

$$\frac{h}{6}f''(x_0) - \frac{h}{6}f''(x_1) = -\frac{2h^2}{18}f^{(3)}(x_0) - \frac{h^2}{18}f^{(3)}(x_1) - \frac{h^3}{36}f^{(4)}(x_0) + O(h^5). \tag{37b}$$

We then subtract (37b) from (36) to eliminate the fourth-order derivative and obtain

$$-\frac{5}{3}\left[\frac{f(x_1) - f(x_0)}{h}\right] + \frac{5}{3}f'(x_0) + \frac{7h}{12}f''(x_0) + \frac{3h}{12}f''(x_1) = -\frac{3h^2}{36}f^{(3)}(x_0) + \frac{h^2}{18}f^{(3)}(x_1) + O(h^5). \tag{38}$$

Dividing (38) by $h/3$ and rearranging the terms give Eq. (34) and hence the proof has been completed.

We now use (34) for the second-order derivative term in (30a) and obtain

$$-\frac{1}{2}\sigma^2_m\left[\frac{7}{4}\frac{\partial^2 U_m((x_m)_0,\tau_{n+1/2})}{\partial x^2_m}+\frac{3}{4}\frac{\partial^2 U_m((x_m)_1,\tau_{n+1/2})}{\partial x^2_m}\right]$$

$$=\frac{5\sigma^2_m}{2}\left[\frac{U_m((x_m)_0,\tau_{n+1/2})-U_m((x_m)_1,\tau_{n+1/2})}{h^2}-\frac{1}{h}\frac{\partial U_m((x_m)_0,\tau_{n+1/2})}{\partial x_m}\right]$$

$$-\frac{\sigma^2_m h}{2}\left[\frac{1}{4}\frac{\partial^3 U_m((x_m)_0,\tau_{n+1/2})}{\partial x^3_m}+\frac{1}{6}\frac{\partial^3 U_m((x_m)_1,\tau_{n+1/2})}{\partial x^3_m}\right]+O(h^4). \quad (39)$$

To evaluate the first-order derivative in (39) at $(x_m)_0$, we use (15), (20) and obtain

$$\frac{\partial U_m((x_m)_0,\tau_{n+1/2})}{\partial x_m}-U_m((x_m)_0,\tau_{n+1/2})=W_m((x_m)_0,\tau_{n+1/2})-U_m((x_m)_0,\tau_{n+1/2})=-K. \quad (40a)$$

To evaluate the third-order derivative term in (39) at $(x_m)_0$, we let $x_m \to 0^+$ in (30b) and discretize $\partial W_m/\partial t$. This gives

$$\frac{\sigma^2_m h}{8}\frac{\partial^3 U_m((x_m)_0,\tau_{n+1/2})}{\partial x^3_m}$$

$$=\frac{h}{4}\left[\frac{W_m((x_m)_0,\tau_{n+1})-W_m((x_m)_0,\tau_n)}{k}-\omega_{n+1/2}Y_m((x_m)_0,\tau_{n+1/2})+(r_m\right.$$

$$\left.-q_{mm})W_m((x_m)_0,\tau_{n+1/2})-\sum_{l\neq m}q_{ml}W_l((x_m)_{j^*|i=0},\tau_{n+1/2})\right]+O(k^2). \quad (40b)$$

Equivalently, the third-order derivative at $(x_m)_1$ is evaluated as follows:

$$\frac{\sigma^2_m h}{12}\frac{\partial^3 U_m((x_m)_1,\tau_{n+1/2})}{\partial x^3_m}$$

$$=\frac{h}{6}\left[\frac{W_m((x_m)_1,\tau_{n+1})-W_m((x_m)_1,\tau_n)}{k}-\omega_{n+1/2}Y_m((x_m)_1,\tau_{n+1/2})+(r_m\right.$$

$$\left.-q_{mm})W_m((x_m)_1,\tau_{n+1/2})-\sum_{l\neq m}q_{ml}W_l((x_m)_{j^*|i=1},\tau_{n+1/2})\right]+O(k^2). \quad (40c)$$

Here, $(x_m)_{j^*|i=0}$ and $(x_m)_{j^*|i=1}$ are the locations in the space for the $l^{th}$ equation corresponding to $(x_m)_0$ and $(x_m)_1$ in the $m^{th}$ equation, respectively, and

$$(\omega_m)_{n+1/2}\equiv\frac{2(s^{n+1}_{f(m)}-s^n_{f(m)})}{k(s^{n+1}_{f(m)}+s^n_{f(m)})}+r_m-\frac{\sigma^2_m}{2}. \quad (41)$$

For the term $Y_m((x_m)_0, \tau_{n+1/2})$ in (30b), we employ a fourth-order approximation (Adam, 1975; Adam, 1976, Liao and Khaliq, 2009; Dremkova and Ehrhardt, 2011) as

$$Y_m((x_m)_0, \tau_{n+1/2})$$
$$= \frac{3}{h}\left[W_m((x_m)_2, \tau_{n+1/2}) - W_m((x_m)_0, \tau_{n+1/2})\right] - 4Y_m((x_m)_1, \tau_{n+1/2}) - Y_m((x_m)_2, \tau_{n+1/2})$$
$$+ O(h^4). \qquad (42)$$

Substituting (40) and (42) into (39), we obtain

$$-\frac{1}{2}\sigma^2{}_m\left[\frac{7}{4}\frac{\partial^2 U_m((x_m)_0, \tau_{n+1/2})}{\partial x_m^2} + \frac{3}{4}\frac{\partial^2 U_m((x_m)_1, \tau_{n+1/2})}{\partial x_m^2}\right]$$
$$= \frac{5\sigma^2{}_m}{2}\left[\frac{U_m((x_m)_0, \tau_{n+1/2}) - U_m((x_m)_1, \tau_{n+1/2})}{h^2}\right] + \frac{5\sigma^2{}_m}{2}\left[\frac{U_m((x_m)_0, \tau_{n+1/2}) - K}{h}\right]$$
$$+ \frac{h}{4}\left[\frac{[U_m((x_m)_0, \tau_{n+1}) - U_m((x_m)_0, \tau_n)]}{k}\right] - \frac{h}{6}\left[\frac{[W_m((x_m)_1, \tau_{n+1}) - W_m((x_m)_1, \tau_n)]}{k}\right]$$
$$- \frac{3}{4}(\omega_m)_{n+1/2}\left(W_m((x_m)_2, \tau_{n+1/2}) - W_m((x_m)_0, \tau_{n+1/2})\right)$$
$$+ \frac{h}{24}(\omega_m)_{n+1/2}\left(14Y_m((x_m)_1, \tau_{n+1/2}) + 3Y_m((x_m)_2, \tau_{n+1/2})\right)$$
$$+ \frac{h(r_m - q_{mm})}{4}U_m((x_m)_0, \tau_{n+1/2}) - \frac{h(r_m - q_{mm})}{4}K - \frac{h(r_m - q_{mm})}{6}W_m((x_m)_1, \tau_{n+1/2})$$
$$- \frac{h}{12}\sum_{l\neq m} q_{ml}\left(3W_l((x_m)_{j^*|i=0}, \tau_{n+1/2}) - 2W_l((x_m)_{j^*|i=1}, \tau_{n+1/2})\right) + O(h^4). \qquad (43)$$

Thus, applying the Crank-Nicholson method in time for (43), we obtain the compact finite difference scheme at $(x_m)_0 = 0$ as

$$\frac{7+h}{4}\left[\frac{(u_m)_0^{n+1} - (u_m)_0^n}{k}\right] + \frac{3}{4}\left[\frac{(u_m)_1^{n+1} - (u_m)_1^n}{k}\right] - \frac{5\sigma^2{}_m}{4h}\left[\frac{(u_m)_1^{n+1} - (u_m)_0^{n+1}}{h} - (u_m)_0^{n+1}\right]$$
$$- \frac{5\sigma^2{}_m}{4h}\left[\frac{(u_m)_1^n - (u_m)_0^n}{h} - (u_m)_0^n\right] - K\left[\frac{5h(r_m - q_{mm})}{4} + \frac{5\sigma^2{}_m}{2h}\right]$$
$$+ \frac{(r_m - q_{mm})}{8}\left[7 + h[(u_m)_0^{n+1} + (u_m)_0^n] + 3[(u_m)_1^{n+1} + (u_m)_1^n]\right] - \frac{h}{6}\left[\frac{(w_m)_1^{n+1} - (w_m)_1^n}{k}\right]$$
$$- \frac{h(r_m - q_{mm})}{12}[(w_m)_1^{n+1} + (w_m)_1^n]$$
$$- \frac{(\omega_m)_{n+1/2}}{8}\left[4[(w_m)_0^{n+1} + (w_m)_0^n] + 3[(w_m)_1^{n+1} + (w_m)_1^n] + 3[(w_m)_2^{n+1} + (w_m)_2^n]\right]$$
$$+ \frac{h(\omega_m)_{n+1/2}}{24}\left[14[(y_m)_1^{n+1} + (y_m)_1^n] + 3[(y_m)_2^{n+1} + (y_m)_2^n]\right]$$

$$-\frac{h}{24}\sum_{l\neq m} q_{ml} \left[3\left((w_l)_{j^*|i=0}^{n+1} + (w_l)_{j^*|i=0}^n\right) - 2\left((w_l)_{j^*|i=1}^{n+1} + (w_l)_{j^*|i=1}^n\right)\right]$$

$$-\frac{1}{8}\sum_{l\neq m} q_{ml}\left[7\left((u_l)_{j^*|i=0}^{n+1} + (u_l)_{j^*|i=0}^n\right) + 3\left((u_l)_{j^*|i=1}^{n+1} + (u_l)_{j^*|i=1}^n\right)\right] = 0, \tag{44}$$

with the truncation error of $O(k^2 + h^4)$. Here, $j^*|i = 0,1$ indicate the values of $j^*$ given at $(x_m)_0$ and $(x_m)_1$, respectively. At each interior grid point, $(x_m)_i = 1,2,\ldots,M-1$, using the compact finite difference scheme (Zhao et al., 2007; Liao and Khaliq, 2009; Cao et al., 2011; Gao and Sun, 2013; Yan et al., 2019) as

$$\frac{1}{12}f''(x_{i-1}) + \frac{10}{12}f''(x_i) + \frac{1}{12}f''(x_{i+1}) = \frac{1}{h^2}[f(x_{i-1}) - 2f(x_i) + f(x_{i+1})] + O(h^4), \tag{45}$$

for (30a)-(30d), we obtain

$$\frac{1}{12}\left[\frac{(u_m)_{i-1}^{n+1} - (u_m)_{i-1}^n}{k}\right] + \frac{10}{12}\left[\frac{(u_m)_i^{n+1} - (u_m)_i^n}{k}\right] + \frac{1}{12}\left[\frac{(u_m)_{i+1}^{n+1} - (u_m)_{i+1}^n}{k}\right]$$

$$-\frac{\sigma^2_m}{4h^2}[(u_m)_{i-1}^{n+1} - 2(u_m)_i^{n+1} + (u_m)_{i+1}^{n+1}] - \frac{\sigma^2_m}{4h^2}[(u_m)_{i-1}^n - 2(u_m)_i^n + (u_m)_{i+1}^n]$$

$$-\frac{(\omega_m)_{n+1/2}}{24}\left[[(w_m)_{i-1}^{n+1} + (w_m)_{i-1}^n] + 10[(w_m)_i^{n+1} + (w_m)_i^n] + [(w_m)_{i+1}^{n+1} + (w_m)_{i+1}^n]\right]$$

$$+\frac{(r_m - q_{mm})}{24}\left[[(u_m)_{i-1}^{n+1} + (u_m)_{i-1}^n] + 10[(u_m)_i^{n+1} + (u_m)_i^n] + [(u_m)_{i+1}^{n+1} + (u_m)_{i+1}^n]\right]$$

$$-\sum_{l\neq m}\frac{q_{ml}}{24}\left[[(u_l)_{j^*-1}^{n+1} + (u_l)_{j^*-1}^n] + 10[(u_l)_{j^*}^{n+1} + (u_l)_{j^*}^n] + [(u_l)_{j^*+1}^{n+1} + (u_l)_{j^*+1}^n]\right]$$

$$= 0, \tag{46a}$$

$$\frac{1}{12}\left[\frac{(w)_{i-1}^{n+1} - (w_m)_{i-1}^n}{k}\right] + \frac{10}{12}\left[\frac{(w_m)_i^{n+1} - (w_m)_i^n}{k}\right] + \frac{1}{12}\left[\frac{(w_m)_{i+1}^{n+1} - (w_m)_{i+1}^n}{k}\right]$$

$$-\frac{\sigma^2_m}{4h^2}[(w_m)_{i-1}^{n+1} - 2(w_m)_i^{n+1} + (w_m)_{i+1}^{n+1}] - \frac{\sigma^2_m}{4h^2}[(w_m)_{i-1}^n - 2(w_m)_i^n + (w_m)_{i+1}^n]$$

$$-\frac{(\omega_m)_{n+1/2}}{2h^2}\left[[(u_m)_{i-1}^{n+1} + (u_m)_{i-1}^n] - 2[(u_m)_i^{n+1} + (u_m)_i^n] + [(u_m)_{i+1}^{n+1} + (u_m)_{i+1}^n]\right]$$

$$+\frac{(r_m - q_{mm})}{24}\left[[(w_m)_{i-1}^{n+1} + (w_m)_{i-1}^n] + 10[(w_m)_i^{n+1} + (w_m)_i^n] + [(w_m)_{i+1}^{n+1} + (w_m)_{i+1}^n]\right]$$

$$-\sum_{l\neq m}\frac{q_{ml}}{24}\left[[(w_l)_{j^*-1}^{n+1} + (w_l)_{j^*-1}^n] + 10[(w_l)_{j^*}^{n+1} + (w_l)_{j^*}^n] + [(w_l)_{j^*+1}^{n+1} + (w_l)_{j^*+1}^n]\right]$$

$$= 0, \tag{46b}$$

$$\frac{1}{12}\left[\frac{(y_m)_{i-1}^{n+1} - (y_m)_{i-1}^n}{k}\right] + \frac{10}{12}\left[\frac{(y_m)_i^{n+1} - (y_m)_i^n}{k}\right] + \frac{1}{12}\left[\frac{(y_m)_{i+1}^{n+1} - (y_m)_{i+1}^n}{k}\right]$$

$$-\frac{\sigma^2_m}{4h^2}[(y_m)^{n+1}_{i-1} - 2(y_m)^{n+1}_i + (y_m)^{n+1}_{i+1}] - \frac{\sigma^2_m}{4h^2}[(y_m)^n_{i-1} - 2(y_m)^n_i + (y_m)^n_{i+1}]$$

$$-\frac{(\omega_m)_{n+1/2}}{2h^2}\Big[[(w_m)^{n+1}_{i-1} + (w_m)^n_{i-1}] - 2[(w_m)^{n+1}_i + (w_m)^n_i] + [(w_m)^{n+1}_{i+1} + (w_m)^n_{i+1}]\Big]$$

$$+\frac{(r_m - q_{mm})}{24}\Big[[(y_m)^{n+1}_{i-1} + (y_m)^n_{i-1}] + 10[(y_m)^{n+1}_i + (y_m)^n_i] + [(y_m)^{n+1}_{i+1} + (y_m)^n_{i+1}]\Big]$$

$$-\sum_{l \neq m} \frac{q_{ml}}{24}\Big[[(y_l)^{n+1}_{j^*-1} + (y_l)^n_{j^*-1}] + 10[(y_l)^{n+1}_{j^*} + (y_l)^n_{j^*}] + [(y_l)^{n+1}_{j^*+1} + (y_l)^n_{j^*+1}]\Big]$$

$$= 0, \tag{46c}$$

$$\frac{1}{12}\left[\frac{(z_m)^{n+1}_{i-1} - (z_m)^n_{i-1}}{k}\right] + \frac{10}{12}\left[\frac{(z_m)^{n+1}_i - (z_m)^n_i}{k}\right] + \frac{1}{12}\left[\frac{(z_m)^{n+1}_{i+1} - (z_m)^n_{i+1}}{k}\right]$$

$$-\frac{\sigma^2_m}{4h^2}[(z_m)^{n+1}_{i-1} - 2(z_m)^{n+1}_i + (z_m)^{n+1}_{i+1}] - \frac{\sigma^2_m}{4h^2}[(z_m)^n_{i-1} - 2(z_m)^n_i + (z_m)^n_{i+1}]$$

$$-\frac{\omega_{n+1/2}}{2h^2}\Big[[(y_m)^{n+1}_{i-1} + (y_m)^n_{i-1}] - 2[(y_m)^{n+1}_i + (y_m)^n_i] + [(y_m)^{n+1}_{i+1} + (y_m)^n_{i+1}]\Big]$$

$$+\frac{(r_m - q_{mm})}{24}\Big[[(z_m)^{n+1}_{i-1} + (z_m)^n_{i-1}] + 10[(z_m)^{n+1}_i + (z_m)^n_i] + [(z_m)^{n+1}_{i+1} + (z_m)^n_{i+1}]\Big]$$

$$-\sum_{l \neq m} \frac{q_{ml}}{24}\Big[[(z_l)^{n+1}_{j^*-1} + (z_l)^n_{j^*-1}] + 10[(z_l)^{n+1}_{j^*} + (z_l)^n_{j^*}] + [(z_l)^{n+1}_{j^*+1} + (z_l)^n_{j^*+1}]\Big]$$

$$= 0, \tag{46d}$$

where $j^*$ represents the location for the $l^{th}$ regime corresponding to $(x_m)_i$, and the truncation error is $O(k^2 + h^4)$. The optimal exercise boundary and the initial and boundary conditions for each regime are calculated as

$$s^{n+1}_{f(m)} = K - (u_m)^{n+1}_0, \quad (w_m)^{n+1}_0 = -s^{n+1}_{f(m)}, \quad (y_m)^{n+1}_0 = -s^{n+1}_{f(m)}, \quad (z_m)^{n+1}_0 = -s^{n+1}_{f(m)}; \tag{47a}$$

$$(u_m)^{n+1}_M = 0, \quad (w_m)^{n+1}_M = 0, \quad (y_m)^{n+1}_M = 0, \quad (z_m)^{n+1}_M = 0; \tag{47b}$$

$$(u_m)^0_i = (w_m)^0_i = (y_m)^0_i = (z_m)^0_i = 0, \quad i = 1,2,\cdots,M. \tag{47c}$$

Let the approximate solutions of the theta, delta decay, and color options for each regime be given as

$$\frac{\partial U_m((x_m)_i, \tau_n)}{\partial \tau} \approx (\Theta_m)^n_i, \quad \frac{\partial W_m((x_m)_i, \tau_n)}{\partial \tau} \approx (K_m)^n_i, \quad \frac{\partial Y_m((x_m)_i, \tau_n)}{\partial \tau} \approx (\Gamma_m)^n_i; \tag{48}$$

respectively. For $n = 1$, we approximate these three Greeks using first-order backward finite differences

$$(\Theta_m)^1_i \approx \frac{(u_m)^1_i - (u_m)^0_i}{k}, \quad (K_m)^1_i \approx \frac{(w_m)^1_i - (w_m)^0_i}{k}, \quad (\Gamma_m)^1_i \approx \frac{(y_m)^1_i - (y)^0_i}{k}. \tag{49a}$$

Subsequently, we use the second-order backward finite difference approximations as

$$(\Theta_m)_i^{n+1} \approx \frac{3(u_m)_i^{n+1} - 4(u_m)_i^n + (u_m)_i^{n-1}}{2k}, \quad (K_m)_i^{n+1} \approx \frac{3(w_m)_i^{n+1} - 4(w_m)_i^n + (w_m)_i^{n-1}}{2k}; \quad (49b)$$

$$(\Gamma_m)_i^{n+1} \approx \frac{3(y_m)_i^{n+1} - 4(y_m)_i^n + (y_m)_i^{n-1}}{2k}. \quad (49c)$$

The initial conditions of the theta, delta decay and color options for each regime are calculated as

$$(\Theta_m)_i^0 = 0, \quad (K_m)_i^0 = 0, \quad (\Gamma_m)_i^0 = 0, \quad i = 0, 1, \cdots, M. \quad (50)$$

### 3.2. Hermite Interpolation

To evaluate $(u_l)_{j^*}^n$, $(w_l)_{j^*}^n$, $(y_l)_{j^*}^n$, and $(z_l)_{j^*}^n$ in (46), we need to consider the relationship between the fixed interval (and the mesh) for the $l^{th}$ regime and the fixed interval (and the mesh) for the $m^{th}$ regime after the logarithmic transformation. If $s_{f(l)}(\tau_n) = s_{f(m)}(\tau_n)$, then the fixed interval for the $l^{th}$ regime overlaps completely with the fixed interval for the $m^{th}$ regime. Hence, $(x_m)_i = (x_l)_j$. If $s_{f(l)}(\tau_n) \neq s_{f(m)}(\tau_n)$, then there are three possible cases as shown in Fig. 1.

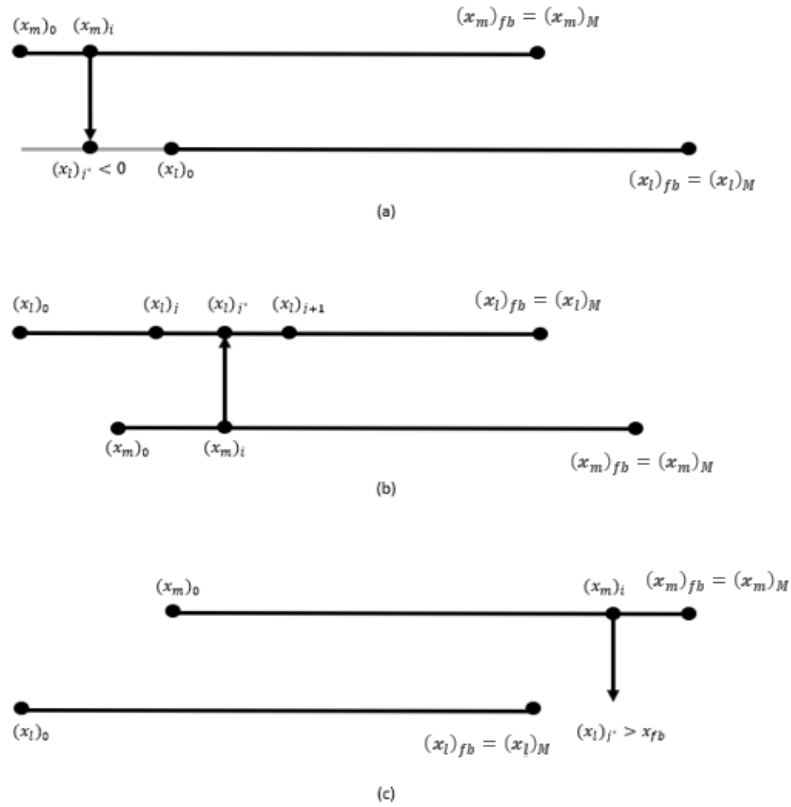

**Fig. 1.** Relationship between the $l^{th}$ and $m^{th}$ intervals and the location of the $(x_m)_i$ in the $l^{th}$ interval.

Fig 1a shows that there exists a possibility for $(x_m)_i$ corresponding to $(x_l)_{j^*} < 0$ in the $l^{th}$ interval. For this case, $(u_l)_i^n = K - s_{f(l)}(\tau_n)e^{(x_l)_{j^*}}$ and $(w_l)_{j^*}^n = (y_l)_{j^*}^n = (z)_{j^*}^n = -s_{f(l)}(\tau_n)e^{(x_l)_{j^*}}$ based on (13)-(16). Fig 1b shows that there exists a possibility for $(x_m)_i$ corresponding to a point in $(0, (x_l)_M)$. For this case, $(u_l)_{j^*}^n, (w_l)_{j^*}^n, (y_l)_{j^*}^n$ and $(z_l)_{j^*}^n$ have to be evaluated using an interpolation based on $(u_l)_j^n, (w_l)_j^n, (y_l)_j^n$ and $(z_l)_j^n$. Here, we employ the Hermite interpolation (Burden et al., 2016) to ensure higher-order accuracy. Fig. 1c shows that there exists a possibility for $(x_m)_i$ corresponding to $(x_l)_{j^*} > (x_l)_M$. For this case, we set $(u_l)_{j^*}^n = (w_l)_{j^*}^n = (y_l)_{j^*}^n = (z_l)_{j^*}^n = 0$. In overall, we can evaluate $(u_l)_{j^*}^n, (w_l)_{j^*}^n, (y_l)_{j^*}^n$ and $(z_l)_{j^*}^n$ based on the following formulas as:

$$(u_l)_{j^*}^n = \begin{cases} K - s_{f(l)}(\tau_n)e^{(x_l)_{j^*}}, & (x_m)_i - \ln\frac{s_{f(l)}(\tau_n)}{s_{f(m)}(\tau_n)} \leq 0; \\ a_c(u_l)_j^n + b_c(u_l)_{j+1}^n + c_c(w_l)_j^n + d_c(w_l)_{j+1}^n, & (x_l)_j \leq (x_m)_i - \ln\frac{s_{f(l)}(\tau_n)}{s_{f(m)}(\tau_n)} \leq (x_l)_{j+1}; \\ 0, & (x_m)_i - \ln\frac{s_{f(l)}(\tau_n)}{s_{f(m)}(\tau_n)} > (x_l)_{fb}, \end{cases} \quad (51a)$$

$$(w_l)_{j^*}^n = \begin{cases} -s_{f(l)}(\tau_n)e^{(x_l)_{j^*}}, & (x_m)_i - \ln\frac{s_{f(l)}(\tau_n)}{s_{f(m)}(\tau_n)} \leq 0; \\ e_c(u_l)_j^n + f_c(u_l)_{j+1}^n + g_c(w_l)_j^n + o_c(w_l)_{j+1}^n, & (x_l)_j \leq (x_m)_i - \ln\frac{s_{f(l)}(\tau_n)}{s_{f(m)}(\tau_n)} \leq (x_l)_{j+1}; \\ 0, & (x_m)_i - \ln\frac{s_{f(l)}(\tau_n)}{s_{f(m)}(\tau_n)} > (x_l)_{fb}, \end{cases} \quad (51b)$$

$$(y_l)_{j^*}^n = \begin{cases} -s_{f(l)}(\tau_n)e^{(x_l)_{j^*}}, & (x_m)_i - \ln\frac{s_{f(l)}(\tau_n)}{s_{f(m)}(\tau_n)} \leq 0; \\ a_c(y_l)_j^n + b_c(y_l)_{j+1}^n + c_c(z_l)_j^n + d_c(z_l)_{j+1}^n, & (x_l)_j \leq (x_m)_i - \ln\frac{s_{f(l)}(\tau_n)}{s_{f(m)}(\tau_n)} \leq (x_l)_{j+1}; \\ 0, & (x_m)_i - \ln\frac{s_{f(l)}(\tau_n)}{s_{f(m)}(\tau_n)} > (x_l)_{fb}, \end{cases} \quad (51c)$$

$$(y_l)_{j^*}^n = \begin{cases} -s_{f(l)}(\tau_n)e^{(x_l)_{j^*}}, & (x_m)_i - \ln\frac{s_{f(l)}(\tau_n)}{s_{f(m)}(\tau_n)} \leq 0; \\ \\ a_c(z_l)_j^n + b_c(z_l)_{j+1}^n + c_c(\dot{z})_j^n + d_c(\dot{z})_{j+1}^n, & (x_l)_j \leq (x_m)_i - \ln\frac{s_{f(l)}(\tau_n)}{s_{f(m)}(\tau_n)} \leq (x_l)_{j+1}; \\ \\ 0, & (x_m)_i - \ln\frac{s_{f(l)}(\tau_n)}{s_{f(m)}(\tau_n)} > (x_l)_{fb}, \quad (51d) \end{cases}$$

where the coefficients are given based on the cubic Hermite Interpolation as follows:

$$a_c = \frac{1}{h^2}\left[1 + \frac{2[(x_l)_{j^*} - (x_l)_j]}{h}\right][(x_l)_{j^*} - (x_l)_{j+1}]^2, \tag{52a}$$

$$b_c = \frac{1}{h^2}\left[1 - \frac{2[(x_l)_{j^*} - (x_l)_{j+1}]}{h}\right][(x_l)_{j^*} - (x_l)_j]^2, \tag{52b}$$

$$c_c = \frac{1}{h^2}[(x_l)_{j^*} - (x_l)_j][(x_l)_{j^*} - (x_l)_{j+1}]^2, \quad d_c = \frac{1}{h^2}[(x_l)_{j^*} - (x_l)_{j+1}][(x_l)_{j^*} - (x_l)_j]^2, \tag{52c}$$

$$e_c = \frac{2}{h^2}\left[1 + \frac{2[(x_l)_{j^*} - (x_l)_j]}{h}\right][(x_l)_j - (x_l)_{j+1}] + \frac{2}{h^3}[(x_l)_{j^*} - (x_l)_{j+1}]^2, \tag{52d}$$

$$f_c = \frac{2}{h^2}\left[1 - \frac{2[(x_l)_{j^*} - (x_l)_{j+1}]}{h}\right][(x_l)_{j^*} - (x_l)_j] + -\frac{2}{h^3}[(x_l)_{j^*} - (x_l)_j]^2, \tag{52e}$$

$$g_c = \frac{2}{h^2}[(x_l)_{j^*} - (x_l)_j][(x_l)_{j^*} - (x_l)_{j+1}] + \frac{1}{h^2}[(x_l)_{j^*} - (x_l)_{j+1}]^2, \tag{52f}$$

$$o_c = \frac{2}{h^2}[(x_l)_{j^*} - (x_l)_{j+1}][(x_l)_{j^*} - (x_l)_j] + \frac{1}{h^2}[(x_l)_{j^*} - (x_l)_j]^2. \tag{52g}$$

Alternatively, one may use the higher accurate quintic Hermite interpolation as

$$(u_l)_{j^*}^n = \begin{cases} K - s_{f(l)}(\tau_n)e^{(x_l)_{j^*}}, & (x_m)_i - \ln\frac{s_{f(l)}(\tau_n)}{s_{f(m)}(\tau_n)} \leq 0; \\ \\ a_q(u_l)_{j-1}^n + b_q(u_l)_j^n + c_q(u_l)_{j+1}^n + d_q(w_l)_{j-1}^n + e_q(w_l)_j^n \\ \qquad + f_q(w_l)_{j+1}^n, \quad (x_l)_j \leq (x_m)_i - \ln\frac{s_{f(l)}(\tau_n)}{s_{f(m)}(\tau_n)} \leq (x_l)_{j+1}; \\ \\ 0, & (x_m)_i - \ln\frac{s_{f(l)}(\tau_n)}{s_{f(m)}(\tau_n)} > (x_l)_{fb}, \quad (53a) \end{cases}$$

$$(w_l)_{j^*}^n = \begin{cases} -s_{f(l)}(\tau_n)e^{(x_l)_{j^*}}, & (x_m)_i - \ln\dfrac{s_{f(l)}(\tau_n)}{s_{f(m)}(\tau_n)} \leq 0; \\ g_q(w_l)_{j-1}^n + o_q(w_l)_j^n + p_q(w_l)_{j+1}^n + q_q(y_l)_{j-1}^n + r_q(y_l)_j^n \\ \qquad + s_q(y_l)_{j+1}^n, & (x_l)_j \leq (x_m)_i - \ln\dfrac{s_{f(l)}(\tau_n)}{s_{f(m)}(\tau_n)} \leq (x_l)_{j+1}; \\ 0, & (x_m)_i - \ln\dfrac{s_{f(l)}(\tau_n)}{s_{f(m)}(\tau_n)} > (x_l)_{fb}, \end{cases} \quad (53b)$$

$$(y_l)_{j^*}^n = \begin{cases} K - s_{f(l)}(\tau_n)e^{(x_l)_{j^*}}, & (x_m)_i - \ln\dfrac{s_{f(l)}(\tau_n)}{s_{f(m)}(\tau_n)} \leq 0; \\ a_q(y_l)_{j-1}^n + b_q(y_l)_j^n + c_q(z_l)_{j+1}^n + d_q(z_l)_{j-1}^n + e_q(z_l)_j^n \\ \qquad + f_q(z_l)_{j+1}^n, & (x_l)_j \leq (x_m)_i - \ln\dfrac{s_{f(l)}(\tau_n)}{s_{f(m)}(\tau_n)} \leq (x_l)_{j+1}; \\ 0, & (x_m)_i - \ln\dfrac{s_{f(l)}(\tau_n)}{s_{f(m)}(\tau_n)} > (x_l)_{fb}, \end{cases} \quad (53c)$$

$$(z_l)_{j^*}^n = \begin{cases} K - s_{f(l)}(\tau_n)e^{(x_l)_{j^*}}, & (x_m)_i - \ln\dfrac{s_{f(l)}(\tau_n)}{s_{f(m)}(\tau_n)} \leq 0; \\ a_q(z_l)_{j-1}^n + b_q(z_l)_j^n + c_q(z_l)_{j+1}^n + d_q(\dot{z}_l)_{j-1}^n + e_q(\dot{z}_l)_j^n \\ \qquad + f_q(\dot{z}_l)_{j+1}^n, & (x_l)_j \leq (x_m)_i - \ln\dfrac{s_{f(l)}(\tau_n)}{s_{f(m)}(\tau_n)} \leq (x_l)_{j+1}; \\ 0, & (x_m)_i - \ln\dfrac{s_{f(l)}(\tau_n)}{s_{f(m)}(\tau_n)} > (x_l)_{fb}, \end{cases} \quad (53d)$$

with the coefficients given as

$$a_q = \left[1 + \frac{3}{h}((x_l)_{j^*} - (x_l)_{j-1})\right] \frac{[((x_l)_{j^*} - (x_l)_j)((x_l)_{j^*} - (x_l)_{j+1})]^2}{4h^4}, \quad (54a)$$

$$b_q = \frac{[((x_l)_{j^*} - (x_l)_{j-1})((x_l)_{j^*} - (x_l)_{j+1})]^2}{h^4}, \quad (54b)$$

$$c_q = \left[1 - \frac{3}{h}((x_l)_{j^*} - (x_l)_{j+1})\right] \frac{[((x_l)_{j^*} - (x_l)_{j-1})((x_l)_{j^*} - (x_l)_j)]^2}{4h^4}, \quad (54c)$$

$$d_q = ((x_l)_{j^*} - (x_l)_{j-1}) \frac{[((x_l)_{j^*} - (x_l)_j)((x_l)_{j^*} - (x_l)_{j+1})]^2}{4h^4}, \quad (54d)$$

$$e_q = ((x_l)_{j^*} - (x_l)_j)\frac{[((x_l)_{j^*} - (x_l)_{j-1})((x_l)_{j^*} - (x_l)_{j+1})]^2}{h^4}, \tag{54e}$$

$$f_q = ((x_l)_{j^*} - (x_l)_{j+1})\frac{[((x_l)_{j^*} - (x_l)_{j-1})((x_l)_{j^*} - (x_l)_j)]^2}{4h^4}, \tag{54f}$$

$$g_q = \left[1 + \frac{3}{h}((x_l)_{j^*} - (x_l)_{j-1})\right]\frac{[2(x_l)_{j^*} - ((x_l)_j + (x_l)_{j+1})][((x_l)_{j^*} - (x_l)_j)((x_l)_{j^*} - (x_l)_{j+1})]}{2h^4}$$
$$+ \frac{3[((x_l)_{j^*} - (x_l)_j)((x_l)_{j^*} - (x_l)_{j+1})]^2}{4h^5}, \tag{54g}$$

$$o_q = \frac{2[2(x_l)_{j^*} - ((x_l)_{j-1} + (x_l)_{j+1})][((x_l)_{j^*} - (x_l)_{j-1})((x_l)_{j^*} - (x_l)_{j+1})]}{h^4}, \tag{54h}$$

$$p_q = \left[1 - \frac{3}{h}((x_l)_{j^*} - (x_l)_{j+1})\right]\frac{[2(x_l)_{j^*} - ((x_l)_{j-1} + (x_l)_j)][((x_l)_{j^*} - (x_l)_{j-1})((x_l)_{j^*} - (x_l)_j)]}{2h^4}$$
$$- \frac{3[((x_l)_{j^*} - (x_l)_{j-1})((x_l)_{j^*} - (x_l)_j)]^2}{4h^5}, \tag{54i}$$

$$q_q = \frac{((x_l)_{j^*} - (x_l)_{j-1})[2(x_l)_{j^*} - ((x_l)_j + (x_l)_{j+1})][((x_l)_{j^*} - (x_l)_j)((x_l)_{j^*} - (x_l)_{j+1})]}{2h^4}$$
$$+ \frac{[((x_l)_{j^*} - (x_l)_j)((x_l)_{j^*} - (x_l)_{j+1})]^2}{4h^4} \tag{54j}$$

$$r_q = \frac{2((x_l)_{j^*} - (x_l)_j)[2(x_l)_{j^*} - ((x_l)_{j-1} + (x_l)_{j+1})][((x_l)_{j^*} - (x_l)_{j-1})((x_l)_{j^*} - (x_l)_{j+1})]}{h^4}$$
$$+ \frac{[((x_l)_{j^*} - (x_l)_{j-1})((x_l)_{j^*} - (x_l)_{j+1})]^2}{h^4} \tag{54k}$$

$$s_q = \frac{2((x_l)_{j^*} - (x_l)_{j+1})[2(x_l)_{j^*} - ((x_l)_{j-1} + (x_l)_j)][((x_l)_{j^*} - (x_l)_{j-1})((x_l)_{j^*} - (x_l)_j)]}{2h^4}$$
$$+ \frac{[((x_l)_{j^*} - (x_l)_{j-1})((x_l)_{j^*} - (x_l)_j)]^2}{4h^4}. \tag{54l}$$

The subscripts "c" and "q" denote the cubic and quintic Hermite interpolations, respectively. It should be pointed out that we have compared with other high-order interpolations when estimating these $(u_l)_{j^*}^n, (w_l)_{j^*}^n, (y_l)_{j^*}^n$ and $(z_l)_{j^*}^n$. Hermite interpolation proves to be accurate, more efficient in handling large state space and very fast in computation. Moreover, it is worth noting that the derivative of $(z_l)_{j^*}^n$, $(\dot{z})_{j^*}^n$ is employed in the cubic and quintic Hermite interpolations. To evaluate $(\dot{z})_{j^*}^n$ with fourth-order accuracy, we obtain it from the speed option PDE by further taking derivative. To carry out an extensive analysis, we further investigate the performance of both cubic and quintic Hermite interpolations in the numerical example section using both Gauss-Seidel and Newton iterative methods.

### 3.3. Stability Analysis

The stability analysis of our numerical schemes is carried out using the matrix form of von Neumann method (see Hirsch (2001) and Liao and Khaliq (2009)). Due to the complex system of the present method, we ignore the coupled regimes $(u_l)_i^n$ and $(w_l)_i^n$, $(y_l)_i^n$ and $(z_l)_i^n$ and let

$$(u_m)_i^n = \lambda_m^n e^{I\beta ih}, \quad (w_m)_i^n = \Phi_m^n e^{I\beta ih}, \quad (y_m)_i^n = \Upsilon_m^n e^{I\beta ih}, \quad (z_m)_i^n = \Psi_m^n e^{I\beta ih}, \quad I = \sqrt{-1}. \tag{55}$$

Denote

$$\mu = \frac{\sigma_m^2 k}{4h^2}, \quad \kappa = (r_m - q_{mm})k, \quad \omega = \left(\frac{2\left(s_{f(m)}^{n+1} - s_{f(m)}^n\right)}{\left(s_{f(m)}^{n+1} + s_{f(m)}^n\right)k} + r_m - \frac{\sigma_m^2}{2}\right)k. \tag{56}$$

Substituting (55), (56) into (46a)-(46d), we obtain

$$\lambda_m^{n+1}\left[1 - \frac{1}{3}\sin^2\left(\frac{\beta h}{2}\right) + 4\mu\sin^2\left(\frac{\beta h}{2}\right) + \frac{\kappa}{2} - \frac{\kappa}{2}\sin^2\left(\frac{\beta h}{2}\right)\right]$$
$$-\lambda_m^n\left[1 - \frac{1}{3}\sin^2\left(\frac{\beta h}{2}\right) - 4\mu\sin^2\left(\frac{\beta h}{2}\right) + \frac{\kappa}{2} - \frac{\kappa}{2}\sin^2\left(\frac{\beta h}{2}\right)\right]$$
$$- \omega\left[\Phi_m^{n+1}\left(\frac{1}{2} - \frac{1}{6}\sin^2\left(\frac{\beta h}{2}\right)\right) + \Phi_m^n\left(\frac{1}{2} - \frac{1}{6}\sin^2\left(\frac{\beta h}{2}\right)\right)\right] = 0, \tag{57a}$$

$$\Phi_m^{n+1}\left[1 - \frac{1}{3}\sin^2\left(\frac{\beta h}{2}\right) + 4\mu\sin^2\left(\frac{\beta h}{2}\right) + \frac{\kappa}{2} - \frac{\kappa}{2}\sin^2\left(\frac{\beta h}{2}\right)7\right]$$
$$- \Phi_m^n\left[1 - \frac{1}{3}\sin^2\left(\frac{\beta h}{2}\right) - 4\mu\sin^2\left(\frac{\beta h}{2}\right) + \frac{\kappa}{2} - \frac{\kappa}{2}\sin^2\left(\frac{\beta h}{2}\right)\right]$$
$$- \omega\left[-4\lambda_m^{n+1}\sin^2\left(\frac{\beta h}{2}\right) - 4\lambda_m^n\sin^2\left(\frac{\beta h}{2}\right)\right] = 0, \tag{57b}$$

$$\Upsilon_m^{n+1}\left[1 - \frac{1}{3}\sin^2\left(\frac{\beta h}{2}\right) + 4\mu\sin^2\left(\frac{\beta h}{2}\right) + \frac{\kappa}{2} - \frac{\kappa}{2}\sin^2\left(\frac{\beta h}{2}\right)\right]$$
$$- \Upsilon_m^n\left[1 - \frac{1}{3}\sin^2\left(\frac{\beta h}{2}\right) - 4\mu\sin^2\left(\frac{\beta h}{2}\right) + \frac{\kappa}{2} - \frac{\kappa}{2}\sin^2\left(\frac{\beta h}{2}\right)\right]$$
$$- \omega\left[-4\Phi_m^{n+1}\sin^2\left(\frac{\beta h}{2}\right) - 4\Phi_m^n\sin^2\left(\frac{\beta h}{2}\right)\right] = 0, \tag{57c}$$

$$\Psi_m^{n+1}\left[1 - \frac{1}{3}\sin^2\left(\frac{\beta h}{2}\right) + 4\mu\sin^2\left(\frac{\beta h}{2}\right) + \frac{\kappa}{2} - \frac{\kappa}{2}\sin^2\left(\frac{\beta h}{2}\right)\right]$$

$$- \Psi_m^n\left[1 - \frac{1}{3}\sin^2\left(\frac{\beta h}{2}\right) - 4\mu\sin^2\left(\frac{\beta h}{2}\right) + \frac{\kappa}{2} - \frac{\kappa}{2}\sin^2\left(\frac{\beta h}{2}\right)\right]$$

$$- \omega\left[-4\Upsilon_m^{n+1}\sin^2\left(\frac{\beta h}{2}\right) - 4\Upsilon_m^n\sin^2\left(\frac{\beta h}{2}\right)\right] = 0, \tag{57d}$$

which can be simplified to

$$p\lambda_m^{n+1} - q\lambda_m^n = r\Phi_m^{n+1} + r\Phi_m^n, \qquad p\Phi_m^{n+1} - q\Phi_m^n = s\lambda_m^{n+1} + s\lambda_m^n, \tag{58a}$$

$$p\Upsilon_m^{n+1} - q\Upsilon_m^n = s\Phi_m^{n+1} + s\Phi_m^n, \qquad p\Psi_m^{n+1} - q\Psi_m^n = s\Upsilon_m^{n+1} + s\Upsilon_m^n, \tag{58b}$$

where

$$p = 1 - \frac{1}{3}\sin^2\left(\frac{\beta h}{2}\right) + 4\mu\sin^2\left(\frac{\beta h}{2}\right) + \frac{\kappa}{2} - \frac{\kappa}{2}\sin^2\left(\frac{\beta h}{2}\right), \qquad r = -\frac{2\omega}{h^2}\sin^2\left(\frac{\beta h}{2}\right), \tag{59a}$$

$$s = \omega\left[\frac{1}{2} - \frac{1}{6}\sin^2\left(\frac{\beta h}{2}\right)\right], \qquad q = 1 - \frac{1}{3}\sin^2\left(\frac{\beta h}{2}\right) - 4\mu\sin^2\left(\frac{\beta h}{2}\right) + \frac{\kappa}{2} - \frac{\kappa}{2}\sin^2\left(\frac{\beta h}{2}\right). \tag{59b}$$

We then obtain a system of equations from (58) and present it in matrix-vector form as

$$\begin{bmatrix} \lambda_m^{n+1} \\ \Phi_m^{n+1} \\ \Upsilon_m^{n+1} \\ \Psi_m^{n+1} \end{bmatrix} = \begin{bmatrix} p & -r & 0 & 0 \\ -s & p & 0 & 0 \\ 0 & -s & p & 0 \\ 0 & 0 & -s & p \end{bmatrix}^{-1} \begin{bmatrix} q & r & 0 & 0 \\ s & q & 0 & 0 \\ 0 & s & q & 0 \\ 0 & 0 & s & q \end{bmatrix} \begin{bmatrix} \lambda_m^n \\ \Phi_m^n \\ \Upsilon_m^n \\ \Psi_m^n \end{bmatrix} =$$

$$\begin{bmatrix} \dfrac{p}{p^2 - sr} & \dfrac{r}{p^2 - sr} & 0 & 0 \\ \dfrac{s}{p^2 - sr} & \dfrac{p}{p^2 - sr} & 0 & 0 \\ \dfrac{s^2}{p(p^2 - sr)} & \dfrac{s}{p^2 - sr} & \dfrac{1}{p} & 0 \\ \dfrac{s^3}{p^2(p^2 - sr)} & \dfrac{s^2}{p(p^2 - sr)} & \dfrac{s}{p^2} & \dfrac{1}{p} \end{bmatrix} \begin{bmatrix} q & r & 0 & 0 \\ s & q & 0 & 0 \\ 0 & s & q & 0 \\ 0 & 0 & s & q \end{bmatrix} \begin{bmatrix} \lambda_m^n \\ \Phi_m^n \\ \Upsilon_m^n \\ \Psi_m^n \end{bmatrix}$$

$$= \begin{bmatrix} \dfrac{pq + rs}{p^2 - sr} & \dfrac{pr + qr}{p^2 - sr} & 0 & 0 \\ \dfrac{ps + qs}{p^2 - sr} & \dfrac{pq + rs}{p^2 - sr} & 0 & 0 \\ \dfrac{ps^2 + qs^2}{p(p^2 - sr)} & \dfrac{pqs + p^2 s}{p(p^2 - sr)} & \dfrac{q}{p} & 0 \\ \dfrac{ps^3 + qs^3}{p^2(p^2 - sr)} & \dfrac{pqs^2 + p^2 s^2}{p^2(p^2 - sr)} & \dfrac{ps + qs}{p^2} & \dfrac{q}{p} \end{bmatrix} \begin{bmatrix} \lambda_m^n \\ \Phi_m^n \\ \Upsilon_m^n \\ \Psi_m^n \end{bmatrix} = A \begin{bmatrix} \lambda_m^n \\ \Phi_m^n \\ \Upsilon_m^n \\ \Psi_m^n \end{bmatrix}. \tag{60}$$

Here, $A$ represents the amplification matrix. To show our numerical method to be unconditionally stable, we need to confirm that the modulus of the eigenvalue of the matrix A is less than or equal to 1 (see

Hirsch (2001) and Liao and Khaliq (2009)). Denoting the eigenvalue of the matrix $A$ as $\varphi$, we obtain the equation below

$$\left[\varphi^2 - 2\varphi \frac{pq+rs}{p^2-sr} + \frac{(pq+rs)^2}{(p^2-sr)^2} - \frac{(pr+qr)(ps+qs)}{(p^2-sr)^2}\right]\left[\left(\frac{q}{p}-\varphi\right)\left(\frac{q}{p}-\varphi\right)\right] = 0. \quad (61)$$

Note that

$$rs = -\frac{2\omega^2}{h^2}\sin^2\left(\frac{\beta h}{2}\right)\left[\frac{1}{2} - \frac{1}{6}\sin^2\left(\frac{\beta h}{2}\right)\right] \leq 0, \quad p \geq q. \quad (62)$$

Since $\mu > 0$, we obtain $p > q$ and hence

$$\left(\frac{q}{p}-\varphi\right)\left(\frac{q}{p}-\varphi\right) = 0, \quad |\varphi_{1,2}| = \left|\frac{q}{p}\right| < 1. \quad (63)$$

Furthermore, we need to obtain $\varphi_{3,4}$ by solving

$$\varphi^2 - 2\varphi \frac{pq+rs}{p^2-sr} + \frac{(pq+rs)^2}{(p^2-sr)^2} - \frac{(pr+qr)(ps+qs)}{(p^2-sr)^2} = 0, \quad (64)$$

which gives

$$\varphi_{3,4} = \frac{(pq+rs) \pm (p+q)\sqrt{rs}}{p^2-rs}. \quad (65)$$

Noticing $\varpi \equiv -rs \geq 0$, we obtain the complex conjugate values of the eigenvalues as

$$\varphi_{3,4} = \frac{(pq-\varpi) \pm (p+q)I\sqrt{\varpi}}{p^2+\varpi}. \quad (66)$$

Thus, we have

$$|\varphi_{3,4}|^2 = \frac{(pq-\varpi)^2 + (p+q)^2\varpi}{(p^2+\varpi)^2} = \frac{(p^2+\varpi)(q^2+\varpi)}{(p^2+\varpi)^2} = \frac{(q^2+\varpi)}{(p^2+\varpi)} \leq 1. \quad (67)$$

Based on the von Neumann analysis, we have proved that our numerical schemes are unconditionally stable.

### 3.4. Computational Procedure with Gauss-Seidel Iteration

The system in (44), (46)-(54) must be solved iteratively. Here, we first present an iterative procedure based on the Gauss-Seidel (GS) iterative method (Kwok, 2011; Chapra, 2012). We initialize $s_{f(m)}^n$, $(u_m)_i^n$, $(w_m)_i^n$, $(y_m)_i^n$, $(z_m)_i^n$, $(\Theta_m)_i^n$, and $(K_m)_i^n$ where $(u_l)_{j^*}^n$, $(w_l)_{j^*}^n$, $(y_l)_{j^*}^n$ and $(z_l)_{j^*}^n$ are calculated based on (51)-(54). We assume that $(u_m)_i^{n+1(\text{It}=0)} = (u_m)_i^n$, $(w_m)_i^{n+1(\text{It}=0)} = (w_m)_i^n$, $(y_m)_i^{n+1(\text{It}=0)} = (y_m)_i^n$, and $(z_m)_i^{n+1(\text{It}=0)} = (z_m)_i^n$, where "It" is the iteration counter. Next, $(u_m)_i^{n+1(\text{It}=1)}$ is computed and $s_{f(m)}^{n+1(\text{It}=1)}$ is obtained from $(u_m)_0^{n+1(\text{It}=1)}$. Subsequently, we compute $(w_m)_i^{n+1(\text{It}=1)}, (y_m)_i^{n+1(\text{It}=1)}, (z_m)_i^{n+1(\text{It}=1)}, (\Theta_m)_i^{n+1(\text{It}=1)}, (K_m)_i^{n+1(\text{It}=1)}$, and $(\Gamma_m)_i^{n+1(\text{It}=1)}$. The iterative process continues until the convergence criterion of both $\max_m \left| s_{f(m)}^{n+1(\text{It}+1)} - s_{f(m)}^{n+1(\text{It})} \right| < \varepsilon$ and $\max_{m,i} \left| (u_m)_i^{n+1(\text{It}+1)} - (u_m)_i^{n+1(\text{It})} \right| < \varepsilon$ is satisfied. An algorithm for obtaining the numerical solutions of the optimal exercise boundary, asset option and the option Greeks in each regime using the GS method is described below.

---

**Algorithm 1.** An algorithm based on the Gauss-Seidel Iteration

1. Initialize $s_{f(m)}^n, (u_m)_i^n, (w_m)_i^n, (y_m)_i^n, (z_m)_i^n, (\Theta_m)_i^n, (K_m)_i^n$, and $(\Gamma_m)_i^n$ for $i = 0,1, \ldots, M$ and $m = 1,2, \ldots, I$
2. **for** $n = 1$ **to** $N$
3. Compute $(u_l)_{j^*}^n$ and $(w_l)_{j^*}^n, (y_l)_{j^*}^n$ and $(z_l)_{j^*}^n$ for $l = 1,2, \ldots, I$ and $l \neq m$ based on (51)-(54)
4. Set $s_{f(m)}^{n+1(\text{It}=0)} = s_{f(m)}^n, (u_m)_i^{n+1(\text{It}=0)} = (u_m)_i^n, (w_m)_i^{n+1(\text{It}=0)} = (w_m)_i^n, (y_m)_i^{n+1(\text{It}=0)} = (y_m)_i^n$, $(z_m)_i^{n+1(\text{It}=0)} = (z_m)_i^n$
5. **while true**
6. Compute $(u_l)_{j^*}^{n+1}, (w_l)_{j^*}^{n+1}, (y_l)_{j^*}^{n+1}$ and $(z_l)_{j^*}^{n+1}$ for $l = 1,2, \ldots, I$ and $l \neq m$ based on (51)-(54)
7. **for** $m = 1$ **to** $I$
8. Compute $(u_m)_i^{n+1(\text{It}+1)}$ and evaluate $s_{f(m)}^{n+1(\text{It}+1)}$ based on (44), (46a) and (47a)
9. Evaluate $(w_m)_i^{n+1(\text{It}+1)}, (y_m)_i^{n+1(\text{It}+1)}, (z_m)_i^{n+1(\text{It}+1)}$ based on (46b)-(46d)
10. **end**
11. **if** $\max_m \left| s_{f(m)}^{n+1(\text{It}+1)} - s_{f(m)}^{n+1(\text{It})} \right| < \varepsilon$ and $\max_{m,i} \left| (u_m)_i^{n+1(\text{It}+1)} - (u_m)_i^{n+1(\text{It})} \right| < \varepsilon$
12. Calculate $(\Theta_m)_i^{n+1(\text{It}+1)}$, and $(K_m)_i^{n+1(\text{It}+1)}$, and $(\Gamma_m)_i^{n+1(\text{It}+1)}$ based on (48), (50)
13. Set $s_{f(m)}^n = s_{f(m)}^{n+1}, (u_m)_i^n = (u_m)_i^{n+1}, (w_m)_i^n = (w_m)_i^{n+1}, (y_m)_i^n = (y)_i^{n+1}$, and $(z_m)_i^n = (z_m)_i^{n+1(\text{It}+1)}$
14. **else**
15. Set $s_{f(m)}^{n+1(\text{It})} = s_{f(m)}^{n+1(\text{It}+1)}, (u_m)_i^{n+1(\text{It})} = (u_m)_i^{n+1(\text{It}+1)}, (w_m)_i^{n+1(\text{It})} = (w_m)_i^{n+1(\text{It}+1)}, (y_m)_i^{n+1(\text{It})} = (y_m)_i^{n+1(\text{It}+1)}$, and $(z_m)_i^{n+1(\text{It})} = (z_m)_i^{n+1(\text{It}+1)}$
16. **end**
17. **end**
18. **end**

---

### 3.5. Computational Procedure with Newton Iterative Method

The Newton method is known to provide quadratic convergence to the solution $F(x) = 0$, and solving our numerical scheme with this method presents an alternative and good choice. Based on (44), (46), we start our iteration in the form

$$F\left(u_m^{n+1(\text{It}=0)}\right) = A_m^u u_m^{n+1(\text{It}=0)} - (b_m^u)^n; \quad F\left(w_m^{n+1(\text{It}=0)}\right) = A_m w_m^{n+1(\text{It}=0)} - (b_m^w)^n, \tag{68a}$$

$$F\left(y_m^{n+1(\text{It}=0)}\right) = A_m y_m^{n+1(\text{It}=0)} - (b_m^y)^n; \quad F\left(z_m^{n+1(\text{It}=0)}\right) = A_m z_m^{n+1(\text{It}=0)} - (b_m^z)^n. \tag{68b}$$

Matrix $A_m^u$ is symmetric, sparse and tridiagonal with constant coefficients likewise $A_m$. The former differs from the latter because of the boundary treatment in (44). Next, we evaluate

$$J\left(u_m^{n+1(\text{It}=0)}\right) \Delta u_m^{n+1(\text{It}=0)} = F\left(u_m^{n+1(\text{It}=0)}\right); \quad J\left(w_m^{n+1(\text{It}=0)}\right) \Delta w_m^{n+1(\text{It}=0)} = F\left(w_m^{n+1(\text{It}=0)}\right), \tag{69a}$$

$$J\left(y_m^{n+1(\text{It}=0)}\right) \Delta y_m^{n+1(\text{It}=0)} = F\left(y_m^{n+1(\text{It}=0)}\right); \quad J\left(z_m^{n+1(\text{It}=0)}\right) \Delta z_m^{n+1(\text{It}=0)} = F\left(z_m^{n+1(\text{It}=0)}\right). \tag{69b}$$

The advantage of our model is that the generated discrete Jacobian matrix is symmetric, sparse and tridiagonal with constant coefficients as one can easily observe from (44), (46). More precisely,

$$J(u_m^n) \equiv A_m^u; \quad J(w_m^n) = J\left(y_m^{n)}\right) = J(z_m^n) \equiv A_m \text{ for all } n \text{ where } m = 1, 2, \dots, I. \tag{70}$$

It presents some nice properties that reduce the cost of computing the Jacobian matrix and enable the use of the Thomas Algorithm for solving $\Delta u_m^{n+1(\text{It}=0)}$, $\Delta u_m^{n+1(\text{It}=0)}$, $\Delta u_m^{n+1(\text{It}=0)}$, and $\Delta u_m^{n+1(\text{It}=0)}$. The next iteration is obtained as follows:

$$u_m^{n+1(\text{It}=1)} = G\left(u_m^{n+1(\text{It}=0)}\right) = u_m^{n+1(\text{It}=0)} - \Delta u_m^{n+1(\text{It}=0)}; \tag{71a}$$

$$s_{f(m)}^{n+1(\text{It}=1)} = K - \left(u_m^{n+1(\text{It}=1)}\right)_{i=0}; \tag{71b}$$

$$w_m^{n+1(\text{It}=1)} = G\left(w_m^{n+1(\text{It}=0)}\right) = w_m^{n+1(\text{It}=0)} - \Delta w_m^{n+1(\text{It}=0)}; \tag{71c}$$

$$y_m^{n+1(\text{It}=1)} = G\left(y_m^{n+1(\text{It}=0)}\right) = y_m^{n+1(\text{It}=0)} - \Delta y_m^{n+1(\text{It}=0)}; \tag{71d}$$

$$z_m^{n+1(\text{It}=1)} = G\left(z_m^{n+1(\text{It}=0)}\right) = z_m^{n+1(\text{It}=0)} - \Delta z_m^{n+1(\text{It}=0)}, \text{ for } m = 1, 2, \dots, I. \tag{71e}$$

The iterative process continues until the convergence criterion of both $\max_m \left|s_{f(m)}^{n+1(\text{It}+1)} - s_{f(m)}^{n+1(\text{It})}\right| < \varepsilon$ and $\max_{m,i} \left|(u_m)_i^{n+1(\text{It}+1)} - (u_m)_i^{n+1(\text{It})}\right| < \varepsilon$ is satisfied. It should be pointed out that to facilitate computation using the Newton method, we adopt the procedure used in the work of Egorova et al. (2016) by treating the coupled regime in the set of the system of PDEs explicitly. Moreover, in the work of Khaliq and Liu (2009), the linear implicit approach was adopted in treating the coupled regime. An algorithm for obtaining the numerical solutions of the optimal exercise boundary, asset option and the option Greeks in each regime using the Newton method is described below.

**Algorithm 2.** An algorithm based on the Newton iteration

1. Initialize $s_{f(m)}^n, \boldsymbol{u}_m^n, \boldsymbol{w}_m^n, \boldsymbol{y}_m^n, \boldsymbol{z}_m^n, \boldsymbol{\Theta}_m^n, \mathbf{K}_m^n$, and $\boldsymbol{\Gamma}_m^n$ for $m = 1, 2, \ldots, I$
2. **for** n = 1 to N
3. Compute $\boldsymbol{u}_l^n, \boldsymbol{w}_l^n, \boldsymbol{y}_l^n, \boldsymbol{z}_l^n$ for $l = 1, 2, \ldots, I$ and $l \neq m$ based on (51)-(54)
4. Set $s_{f(m)}^{n+1(It=0)} = s_{f(m)}^n, \boldsymbol{u}_m^{n+1(It=0)} = \boldsymbol{u}_m^n, \boldsymbol{w}_m^{n+1(It=0)} = \boldsymbol{w}_m^n, \boldsymbol{y}_m^{n+1(It=0)} = \boldsymbol{y}_m^n, \boldsymbol{z}_m^{n+1(It=0)} = \boldsymbol{z}_m^n$
5. **while** true
6. **for** m = 1 to I
7. Compute $F\left(\boldsymbol{u}_m^{n+1(It=0)}\right)$. Obtain $\Delta \boldsymbol{u}_m^{n+1(It=0)}$ using the Thomas Algorithm with $J(\boldsymbol{u}_m^n) \equiv A_m^u$ based on (70). Compute $\boldsymbol{u}_m^{n+1(It=1)}$ based on (71) and evaluate $s_{f(m)}^{n+1(It=1)}$ from $\left(u_m^{n+1(It=1)}\right)_{i=0}$ based on (47a)
8. **end**
9. **if** $\max_m \left|s_{f(m)}^{n+1(It+1)} - s_{f(m)}^{n+1(It)}\right| < \varepsilon$ and $\max_{m,i} \left|\boldsymbol{u}_m^{n+1(It+1)} - \boldsymbol{u}_m^{n+1(It)}\right| < \varepsilon$
10. Calculate $\boldsymbol{\Theta}_m^{n+1}$ based on (48)-(50)
11. Set $s_{f(m)}^n = s_{f(m)}^{n+1}, \boldsymbol{u}_m^n = \boldsymbol{u}_m^{n+1}$
12. **else**
13. Set $s_{f(m)}^{n+1(It)} = s_{f(m)}^{n+1(It+1)}, \boldsymbol{u}_m^{n+1(It)} = \boldsymbol{u}_m^{n+1(It+1)}$
14. **end**
15. **end**
16. **while** true
17. Compute $\boldsymbol{w}_m^{n+1(It+1)}$ in the same manner as $\boldsymbol{u}_m^{n+1(It+1)}$ based on (46), (70), (71)
18. Calculate $\mathbf{K}_m^{n+1}$ based on (48)-(50)
19. **end**
20. **while** true
21. Compute $\boldsymbol{y}_m^{n+1(It+1)}$ in the same manner as $\boldsymbol{u}_m^{n+1(It+1)}$ based on (46), (70), (71)
22. Calculate $\boldsymbol{\Gamma}_m^{n+1}$ based on (48)-(50)
23. **end**
24. **while** true
25. Compute $\boldsymbol{z}_m^{n+1(It+1)}$ in the same manner as $\boldsymbol{u}_m^{n+1(It+1)}$ based on (46), (70), (71)
26. **end**
27. **end**

## 4. Numerical Experiments

To test the accuracy and applicability of the present scheme, we consider the American put options pricing problems with two regimes, four regimes, eight regimes, and sixteen regimes, respectively. The numerical code was written with MATLAB 2019a on Intel Core i5-3317U CPU 1.70GHz 64-bit ASUS Laptop. The numerical procedures were carried out on the mesh with a uniform grid size.

## 4.1. Numerical Examples: Two Regimes

We first consider the American put options with two regimes. We label our present method as "FF-CS1, FF-CS2, FF-CS3 and FF-CS4 which we denote as the front fixing-compact scheme with cubic Hermite interpolation and GS iteration, with quintic Hermite interpolation and GS iteration, with cubic Hermite interpolation and Newton Iteration, and with quintic Hermite interpolation and Newton Iteration, respectively. We further compare them with MTree (Liu, 2010), IMS1, IMS2 (Khaliq and Liu, 2009), MOL (Chiarella et al., 2016), RBF-FD (Li et al., 2018), FF-expl (Egorova et al., 2016), ETD-CN (Khaliq et al., 2013), Iterated Optimal Stopping and Local Optimal Iteration (Babbin et al., 2011) as listed in Tables 1-4, and 7. The option Greeks results were also listed in Tables 5 and 6.

**Example 1:** We consider a switching regime problem with the strike price chosen to be $K = 9$ at the expiration time $T = 1$. In our computation, we chose the interval $0 \leq x_m \leq 3$ with the grid size $h = 0.1$, 0.05, and 0.01 for FF-CS1 and FF-CS2 and $h = 0.01$ for FF-CS3 and FF-CS4. The time step $k$ was determined using $k = h^2$. The parameters were given as

$$Q = \begin{bmatrix} -6 & 6 \\ 9 & -9 \end{bmatrix}, \quad r = \begin{bmatrix} 0.10 \\ 0.05 \end{bmatrix}, \quad \sigma = \begin{bmatrix} 0.80 \\ 0.30 \end{bmatrix}, \quad \varepsilon = 10^{-8} \tag{72}$$

Figs. 2 and 3 show the profiles of the option prices, Greek parameters, and optimal exercise boundaries for the two-regime case. From Tables 1-3, one can easily observe that the data obtained based on FF-CS1 and FF-CS2, FF-CS3 and FF-CS4, respectively, when $h = 0.01$ are the same as those obtained from MOL, MTree and RBF-FD up to 5 digits in most cases. In particular, FF-CS3 and FF-CS4 are more than five times faster than FF-CS1 and FF-CS2. Moreover, Chiarella et al. (2016) pointed out that data obtained from MTree data was used as the benchmark in the work of Khaliq and Liu (2009). Generally, our data slightly decreases in direct proportion with $h$.

**Example 2:** We investigate the performance of our method as compared with MOL when there is no jump between regimes (Chiarella et al., 2016; Meyer and van der Hoek, 1997). We use the same data provided in the first example. The grid size and generator matrix were chosen to be $h = 0.01$ and

$$Q = \begin{bmatrix} 0 & 0 \\ 0 & 0 \end{bmatrix}, \tag{73}$$

respectively. The obtained result was listed in Table 8. It can be seen from Table 8 that the data obtained based on FF-CS1 and FF-CS2 are virtually the same. This is because there was no jump between different states.

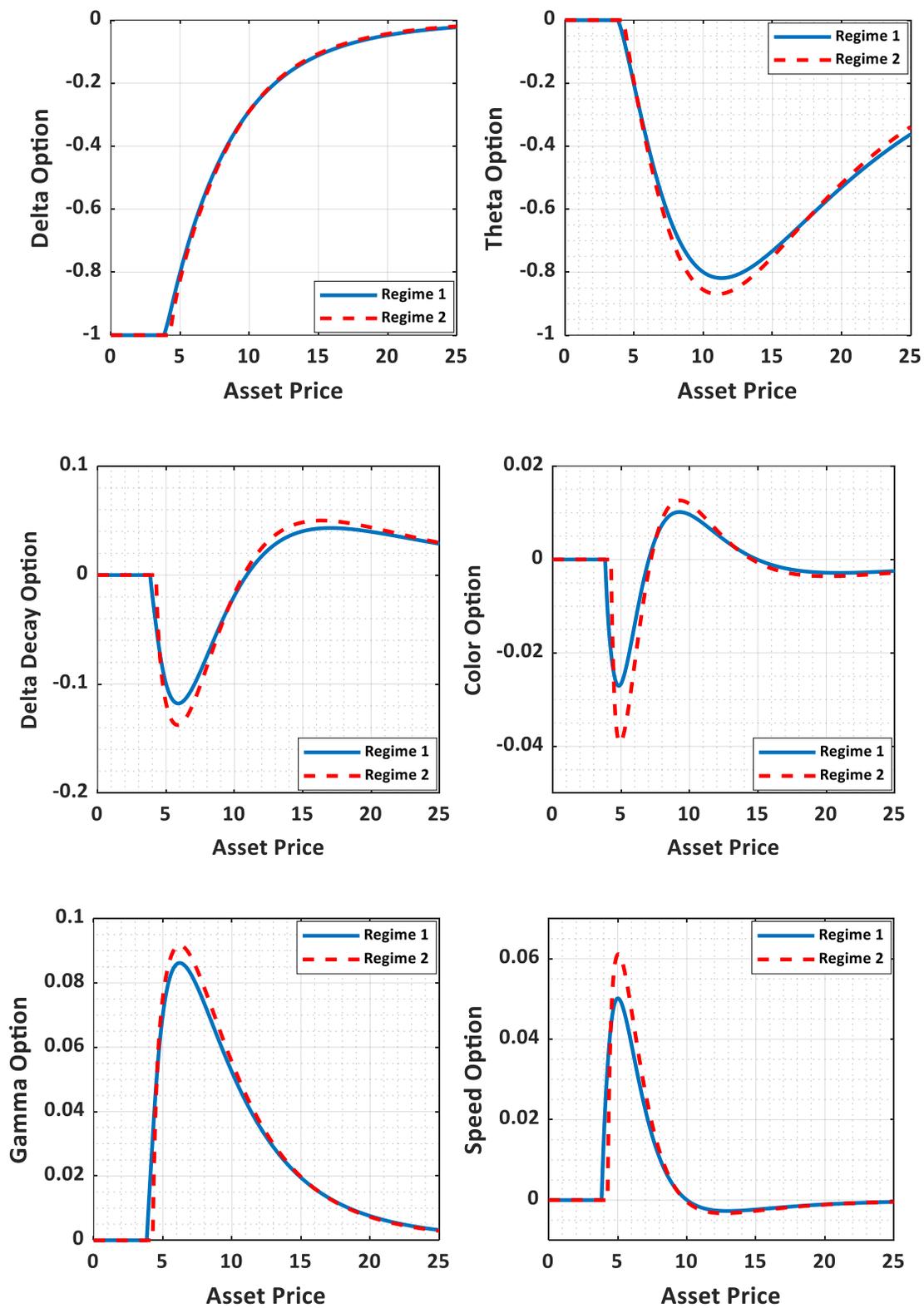

**Fig. 2.** Option Greeks for the two-regime case when $\tau = T$ (example 1).

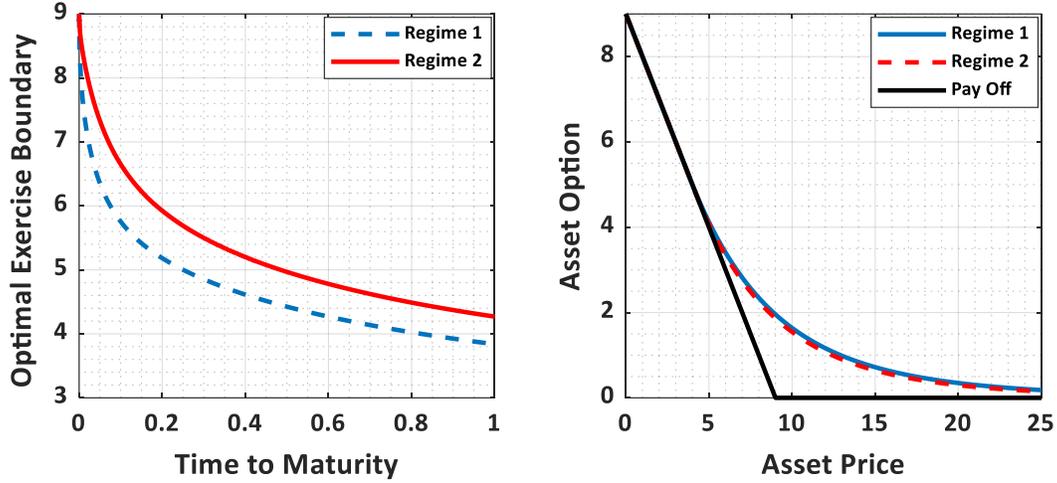

**Fig. 3.** Asset options and optimal exercise boundaries for the two-regime case when $\tau = T$ (example 1).

**Example 3**: We compare our results with Iterated Optimal Stopping and Local Optimal Iteration (Babbin et al., 2011). The strike price was chosen to be $K = 10$ at the expiration time $T = 1$. The grid size was chosen to be $h = 0.01$. The parameters were given as

$$Q = \begin{bmatrix} -3 & 3 \\ 2 & -2 \end{bmatrix}, \quad r = \begin{bmatrix} 0.05 \\ 0.05 \end{bmatrix}, \quad \sigma = \begin{bmatrix} 0.3 \\ 0.4 \end{bmatrix}. \tag{74}$$

Like other given examples, the convergence criterion $\varepsilon = 10^{-8}$ was chosen and the time step $k$ was determined using $k = h^2$. In Table 7, we compared the result with RBF-FD, IOS, and LOP methods. The gamma and speed plots from this example were shown in Fig. 4. To check the accuracy of our present method, we calculated the convergence rate from the asset option in regime 1. To obtain the convergence rate of our numerical scheme, we defined the maximum error using the notation

**Table 1.** Comparison of American put option price in regime 1 for example 1.

| S | MTree | IMS1 | IMS2 | MOL | FF-CS1 | | | FF-CS2 | | | FF-CS3 | FF-CS4 |
|---|---|---|---|---|---|---|---|---|---|---|---|---|
|   |   |   |   |   | $h = 0.1$ | 0.05 | 0.01 | 0.1 | 0.05 | 0.01 | 0.01 | 0.01 |
| 3.5 | 5.5000 | 5.5001 | 5.5001 | 5.5000 | 5.5000 | 5.5000 | 5.5000 | 5.5000 | 5.5000 | 5.0000 | 5.5000 | 5.0000 |
| 4.0 | 5.0066 | 5.0067 | 5.0066 | 5.0033 | 5.0068 | 5.0035 | 5.0033 | 5.5069 | 5.0035 | 5.0033 | 5.0033 | 5.0033 |
| 4.5 | 4.5432 | 4.5486 | 4.5482 | 4.5433 | 4.5475 | 4.5442 | 4.5433 | 4.5475 | 4.5442 | 4.5433 | 4.5433 | 4.5433 |
| 6.0 | 3.4144 | 3.4198 | 3.4184 | 3.4143 | 3.4189 | 3.4142 | 3.4143 | 3.4190 | 3.4143 | 3.4143 | 3.4141 | 3.4141 |
| 7.5 | 2.5844 | 2.5877 | 2.5867 | 2.5842 | 2.5873 | 2.5854 | 2.5842 | 2.5874 | 2.5854 | 2.5842 | 2.5840 | 2.5840 |
| 8.5 | 2.1560 | 2.1598 | 2.1574 | 2.1559 | 2.1539 | 2.1553 | 2.1559 | 2.1540 | 2.1553 | 2.1559 | 2.1556 | 2.1556 |
| 9.0 | 1.9722 | 1.9756 | 1.9731 | 1.9720 | 1.9759 | 1.9722 | 1.9720 | 1.9760 | 1.9723 | 1.9720 | 1.9717 | 1.9717 |
| 9.5 | 1.8058 | 1.8090 | 1.8064 | 1.8056 | 1.8043 | 1.8062 | 1.8056 | 1.8044 | 1.8062 | 1.8056 | 1.8054 | 1.8054 |
| 10.5 | 1.5186 | 1.5214 | 1.5187 | 1.5185 | 1.5170 | 1.5190 | 1.5185 | 1.5170 | 1.5190 | 1.5185 | 1.5183 | 1.5183 |
| 12.0 | 1.1803 | 1.1827 | 1.1799 | 1.1803 | 1.1833 | 1.1802 | 1.1803 | 1.1833 | 1.1802 | 1.1803 | 1.1801 | 1.1801 |

**Table 2.** Comparison of American put option price in regime 2 for example 1.

| S | MTree | IMS1 | IMS2 | MOL | FF-CS1 | | | FF-CS2 | | | FF-CS3 | FF-CS4 |
|---|---|---|---|---|---|---|---|---|---|---|---|---|
| | | | | | $h=0.1$ | 0.05 | 0.01 | 0.1 | 0.05 | 0.01 | 0.01 | 0.01 |
| 3.5 | 5.5000 | 5.5012 | 5.5012 | 5.5000 | 5.5000 | 5.5000 | 5.5000 | 5.5000 | 5.5000 | 5.5000 | 5.5000 | 5.5000 |
| 4.0 | 5.0000 | 5.0016 | 5.0016 | 5.0000 | 5.0000 | 5.0000 | 5.0000 | 5.0000 | 5.0000 | 5.0000 | 5.0000 | 5.0000 |
| 4.5 | 4.5117 | 4.5194 | 4.5190 | 4.5119 | 4.5183 | 4.5129 | 4.5119 | 4.5184 | 4.5129 | 4.5119 | 4.5119 | 4.5119 |
| 6.0 | 3.3503 | 3.3565 | 3.3550 | 3.3507 | 3.3552 | 3.3508 | 3.3507 | 3.3553 | 3.3508 | 3.3507 | 3.3504 | 3.3504 |
| 7.5 | 2.5028 | 2.5078 | 2.5056 | 2.5033 | 2.5070 | 2.5044 | 2.5033 | 2.5071 | 2.5045 | 2.5033 | 2.5030 | 2.5030 |
| 8.5 | 2.0678 | 2.0722 | 2.0695 | 2.0683 | 2.0677 | 2.0683 | 2.0683 | 2.0679 | 2.0684 | 2.0683 | 2.0681 | 2.0681 |
| 9.0 | 1.8819 | 1.8860 | 1.8832 | 1.8825 | 1.8864 | 1.8822 | 1.8825 | 1.8864 | 1.8820 | 1.8825 | 1.8822 | 1.8822 |
| 9.5 | 1.7143 | 1.7181 | 1.7153 | 1.7149 | 1.7116 | 1.7149 | 1.7149 | 1.7116 | 1.7149 | 1.7149 | 1.7146 | 1.7146 |
| 10.5 | 1.4267 | 1.4301 | 1.4272 | 1.4273 | 1.4240 | 1.4273 | 1.4273 | 1.4239 | 1.4274 | 1.4273 | 1.4271 | 1.4271 |
| 12.0 | 1.0916 | 1.0945 | 1.0916 | 1.0923 | 1.0948 | 1.0927 | 1.0923 | 1.0948 | 1.0927 | 1.0923 | 1.0921 | 1.0921 |

**Table 3.** Further Comparison of American put option price for example 1.

| S | RBF-FD | | ETD-CN | | FF-expl | | FF-CS2 ($h=0.01$) | |
|---|---|---|---|---|---|---|---|---|
| | Regime 1 | Regime 2 | Regime 1 | Regime 2 | Regime 1 | Regime 2 | Regime 1 | Regime 2 |
| 9.0 | 1.9718 | 1.8825 | 1.9756 | 1.8859 | 1.9713 | 1.8817 | 1.9720 | 1.8825 |
| 10.5 | 1.5185 | 1.4274 | 1.5213 | 1.4301 | 1.5177 | 1.4265 | 1.5185 | 1.4273 |
| 12.0 | 1.1803 | 1.0924 | 1.1825 | 1.0945 | 1.1796 | 1.0915 | 1.1803 | 1.0923 |

**Table 4.** Comparing FF-CS3, and FF-CS4 up to sixteen digits at strike price for example 1.

| Strike price | FF-CS3 | | FF-CS4 | |
|---|---|---|---|---|
| | Regime 1 | Regime 2 | Regime 1 | Regime 2 |
| 9.0 | 1.971738757801249 | 1.882203676543793 | 1.971733636374602 | 1.882198321043946 |

**Table 5.** American put option Greeks for the two-regime case for example 1 with FF-CS2.

| | Delta | | Gamma | | Speed | |
|---|---|---|---|---|---|---|
| S | Regime 1 | Regime 2 | Regime 1 | Regime 2 | Regime 1 | Regime 2 |
| 3.5 | -1.0000 | -1.0000 | 0.0000 | 0.0000 | 0.0000 | 0.0000 |
| 4.0 | -0.9652 | -1.0000 | 0.0164 | 0.0000 | 0.0171 | 0.0000 |
| 4.5 | -0.8749 | -0.9171 | 0.0508 | 0.0497 | 0.0438 | 0.0470 |
| 6.0 | -0.6426 | -0.6571 | 0.0851 | 0.0905 | 0.0381 | 0.0462 |
| 9.5 | -0.3165 | -0.3181 | 0.0560 | 0.0594 | 0.0015 | 0.0013 |
| 12.0 | -0.1945 | -0.1913 | 0.0347 | 0.0361 | -0.0025 | -0.0031 |

**Table 6.** American put option Greeks for the two-regime case for example 1 with FF-CS2.

| | Theta | | Delta-Decay | | Color | |
|---|---|---|---|---|---|---|
| S | Regime 1 | Regime 2 | Regime 1 | Regime 2 | Regime 1 | Regime 2 |
| 3.5 | 0.0000 | 0.0000 | 0.0000 | 0.0000 | 0.0000 | 0.0000 |
| 4.0 | -0.0300 | 0.0000 | -0.0211 | 0.0000 | -0.0086 | 0.0000 |
| 4.5 | -0.1200 | -0.0850 | -0.0690 | -0.0722 | -0.0229 | -0.0295 |
| 6.0 | -0.4083 | -0.4279 | -0.1160 | -0.1358 | -0.0125 | -0.0206 |
| 9.5 | -0.7904 | -0.8467 | -0.0310 | -0.0317 | 0.0108 | 0.0132 |
| 12.0 | 0.8248 | -0.8700 | 0.0169 | 0.0240 | 0.0061 | 0.0069 |

**Table 7.** Comparison of American put option price with IOS and LOI in Regime 1 for example 3.

| S | IOS | LOI | RBF-FD | FF-CS2 | FF-CS4 |
|---|---|---|---|---|---|
| | Maximum Refinement | | | $h = 0.01$ | |
| 10.00 | 1.1747961 | 1.1747960 | 1.1756722 | 1.1750372 | 1.1751356 |

**Table 8.** Comparison of American put option price with no jump between regimes for example 2.

| S | MOL | | FF-CS1 | | FF-CS2 | |
|---|---|---|---|---|---|---|
| | Regime 1 | Regime 2 | Regime 1 | Regime 2 | Regime 1 | Regime 2 |
| 6.00 | 3.666762424 | 3.000000000 | 3.666746420 | 3.000000000 | 3.666746420 | 3.000000000 |
| 9.00 | 2.375385605 | 0.888311178 | 2.375408073 | 0.888393716 | 2.375408073 | 0.888393716 |
| 12.00 | 1.604853957 | 0.203543056 | 1.604912489 | 0.204583983 | 1.604912489 | 0.203637945 |

**Table 9.** The maximum errors and convergence rates for regime 1 for example 1.

| $h$ | maximum error | | convergence rate | |
|---|---|---|---|---|
| | (FF-CS1) | (FF-CS2) | (FF-CS1) | (FF-CS2) |
| $2 \times 10^{-1}$ | | | | |
| $1 \times 10^{-1}$ | $5.344 \times 10^{-2}$ | $5.128 \times 10^{-2}$ | | |
| $5 \times 10^{-2}$ | $6.269 \times 10^{-3}$ | $6.196 \times 10^{-3}$ | 3.09 | 3.05 |
| $2.5 \times 10^{-2}$ | $6.329 \times 10^{-4}$ | $6.806 \times 10^{-4}$ | 3.31 | 3.19 |

**Table 10.** Average CPU time(s) per each time step for the two-regime example.

| $h$ | CPU Time(s) | | | |
|---|---|---|---|---|
| | FF-CS1 | FF-CS2 | FF-CS3 | FF-CS4 |
| 0.1 | 0.182 | 0.209 | 0.058 | 0.061 |
| 0.05 | 0.311 | 0.316 | 0.084 | 0.078 |
| 0.01 | 3.697 | 4.167 | 0.768 | 0.771 |

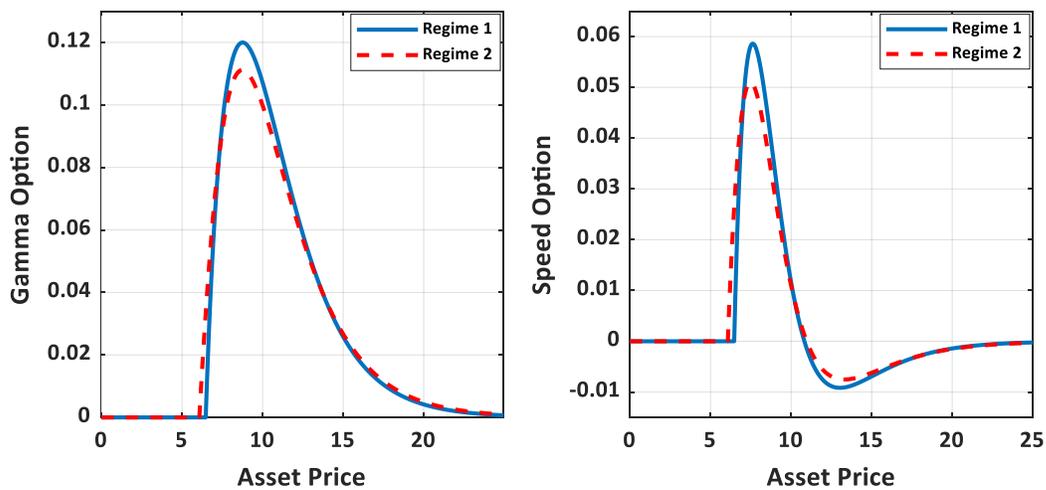

**Fig. 4.** Gamma and speed options for the two-regime case when $\tau = T$ (example 3).

$$E(h, k) = \max_{0 \leq i \leq M} |(u_1)_i^n(h, k) - (u_1)_i^n(h/2, k/4)|, \tag{75a}$$

$$E(h/2, k/4) = \max_{0 \leq i \leq M} |(u_1)_i^n(h/2, k/4) - (u_1)_i^n(h/4, k/16)|, \tag{75b}$$

where $k = h^2$. $(u_1)_i^n(h, k), (u_1)_i^n(h/2, k/4)$, and $(u_1)_i^n(h/4, k/16)$ are the numerical solutions from regime 1 obtained based on $h$ and $k$, $h/2$ and $k/4$, and $h/4$ and $k/16$, respectively. As such, the convergence rate was evaluated using the following equation as:

$$Rate = \log_2 \frac{E(h, k)}{E(h/2, k/4)}. \tag{76}$$

Table 9 lists the maximum errors and convergence rates of FF-CS1 and FF-CS2 obtained based on $h = 0.2, 0.1, 0.05, 0.025, 0.0125$. It can be seen from Table 9 that the convergence rate is above 3.0, indicating that our present method provides a more accurate solution than the existing methods do. Besides, the computational speed of FF-CS1, FF-CS2, FF-CS3, and FF-CS4 is very fast as seen from Table 10.

### 4.2. Numerical Examples beyond Two Regimes

Commonly, previous works of literature have limited the regime-switching analysis to two and four regimes. Moreover, Chiarella et al. (2016) and Khaliq and Liu (2008) pointed out that the method proposed by Buffington and Elliot (2002) cannot be extended beyond two regimes. To show that our method can compute a large finite state space, we wrote a sequence of MATLAB function files and used it to write a few lines of code that can take any number of finite state spaces. We then considered the American put options pricing problems with four, eight, and sixteen regimes, respectively. The strike price and expiration time were chosen to be $K = 9$ and $T = 1$, respectively. In our computation, we chose the interval $0 \leq x_m \leq 3$ where the grid size $h = 10^{-2}$ and $k = 10^{-4}$. The four-regime example was computed with the convergent criterion of $\varepsilon = 10^{-8}$ while eight- and sixteen-regime examples were computed with $\varepsilon = 10^{-7}$ and $\varepsilon = 10^{-8}$ for FF-CS2 and FF-CS4, respectively. The parameters are given in (77)-(79), respectively.

**Four-regime example**:

$$Q = \begin{bmatrix} -1 & 1/3 & 1/3 & 1/3 \\ 1/3 & -1 & 1/3 & 1/3 \\ 1/3 & 1/3 & -1 & 1/3 \\ 1/3 & 1/3 & 1/3 & -1 \end{bmatrix}, \quad r = \begin{bmatrix} 0.02 \\ 0.10 \\ 0.06 \\ 0.15 \end{bmatrix}, \quad \sigma = \begin{bmatrix} 0.90 \\ 0.50 \\ 0.70 \\ 0.20 \end{bmatrix}. \tag{77}$$

**Eight-regime example**:

$$Q = \begin{bmatrix} -1 & 0.2 & 0.2 & 0.2 & 0.1 & 0.1 & 0.1 & 0.1 \\ 0.2 & -1 & 0.1 & 0.1 & 0.1 & 0.2 & 0.2 & 0.1 \\ 0.2 & 0.1 & -1 & 0.1 & 0.2 & 0.1 & 0.1 & 0.2 \\ 0.2 & 0.1 & 0.2 & -1 & 0.2 & 0.1 & 0.1 & 0.1 \\ 0.1 & 0.2 & 0.1 & 0.1 & -1 & 0.2 & 0.1 & 0.2 \\ 0.2 & 0.2 & 0.2 & 0.1 & 0.1 & -1 & 0.1 & 0.1 \\ 0.1 & 0.1 & 0.2 & 0.2 & 0.2 & 0.1 & -1 & 0.1 \\ 0.1 & 0.1 & 0.1 & 0.2 & 0.1 & 0.2 & 0.2 & -1 \end{bmatrix}, \quad r = \begin{bmatrix} 0.03 \\ 0.15 \\ 0.20 \\ 0.09 \\ 0.05 \\ 0.12 \\ 0.15 \\ 0.18 \end{bmatrix}, \quad \sigma = \begin{bmatrix} 0.80 \\ 0.40 \\ 0.50 \\ 0.70 \\ 0.45 \\ 0.38 \\ 0.30 \\ 0.25 \end{bmatrix}. \tag{78}$$

**Sixteen-regime example:**

$$r = [\,0.04\ 0.15\ 0.03\ 0.30\ 0.13\ 0.12\ 0.10\ 0.18\ 0.08\ 0.25\ 0.06\ 0.20\ 0.21\ 0.07\ 0.12\ 0.19\,],$$

$$\sigma = [\,0.07\ 0.30\ 0.90\ 0.80\ 0.25\ 0.15\ 0.12\ 0.28\ 0.85\ 0.35\ 0.39\ 0.72\ 0.45\ 0.18\ 0.20\ 0.25\,],$$

$$Q = \begin{bmatrix} -3 & 0.2 & 0.2 & 0.2 & 0.2 & \cdots & 0.2 & 0.2 & 0.2 & 0.2 & 0.2 \\ 0.2 & -3 & 0.2 & 0.2 & 0.2 & \cdots & 0.2 & 0.2 & 0.2 & 0.2 & 0.2 \\ 0.2 & 0.2 & -3 & 0.2 & 0.2 & \cdots & 0.2 & 0.2 & 0.2 & 0.2 & 0.2 \\ 0.2 & 0.2 & 0.2 & -3 & 0.2 & \cdots & 0.2 & 0.2 & 0.2 & 0.2 & 0.2 \\ 0.2 & 0.2 & 0.2 & 0.2 & -3 & \cdots & 0.2 & 0.2 & 0.2 & 0.2 & 0.2 \\ \vdots & \vdots & \vdots & \vdots & \vdots & \ddots & \vdots & \vdots & \vdots & \vdots & \vdots \\ 0.2 & 0.2 & 0.2 & 0.2 & 0.2 & \cdots & 0.2 & -3 & 0.2 & 0.2 & 0.2 \\ 0.2 & 0.2 & 0.2 & 0.2 & 0.2 & \cdots & 0.2 & 0.2 & -3 & 0.2 & 0.2 \\ 0.2 & 0.2 & 0.2 & 0.2 & 0.2 & \cdots & 0.2 & 0.2 & 0.2 & -3 & 0.2 \\ 0.2 & 0.2 & 0.2 & 0.2 & 0.2 & \cdots & 0.2 & 0.2 & 0.2 & 0.2 & -3 \end{bmatrix}. \tag{79}$$

Figs. 5-8 plot the profiles of the option prices, Greek parameters, and optimal exercise boundaries for the four, eight, and sixteen regimes. Tables 11-14 list the option prices and Greeks of the four, eight and sixteen regimes using the asset values in the interval of $3.5 \leq S \leq 12$.

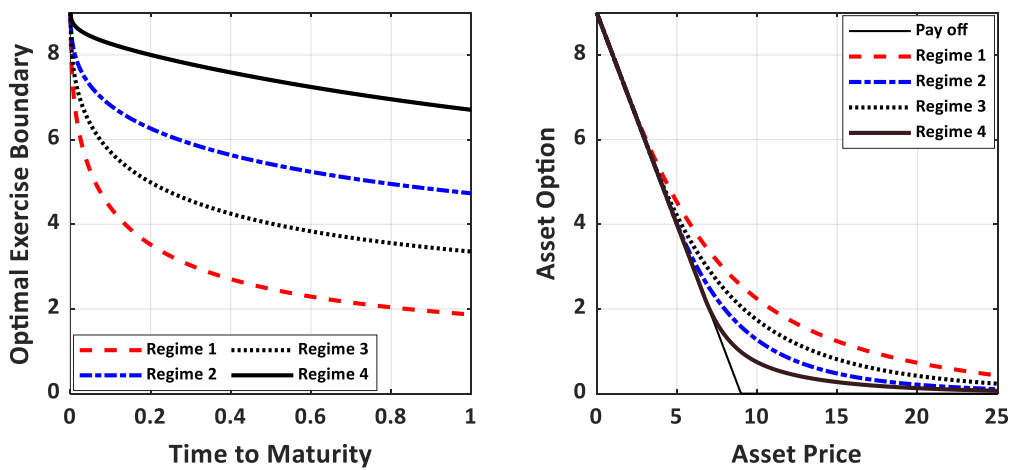

**Fig. 5.** Asset options and optimal exercise boundaries for the four-regime case when $\tau = T$.

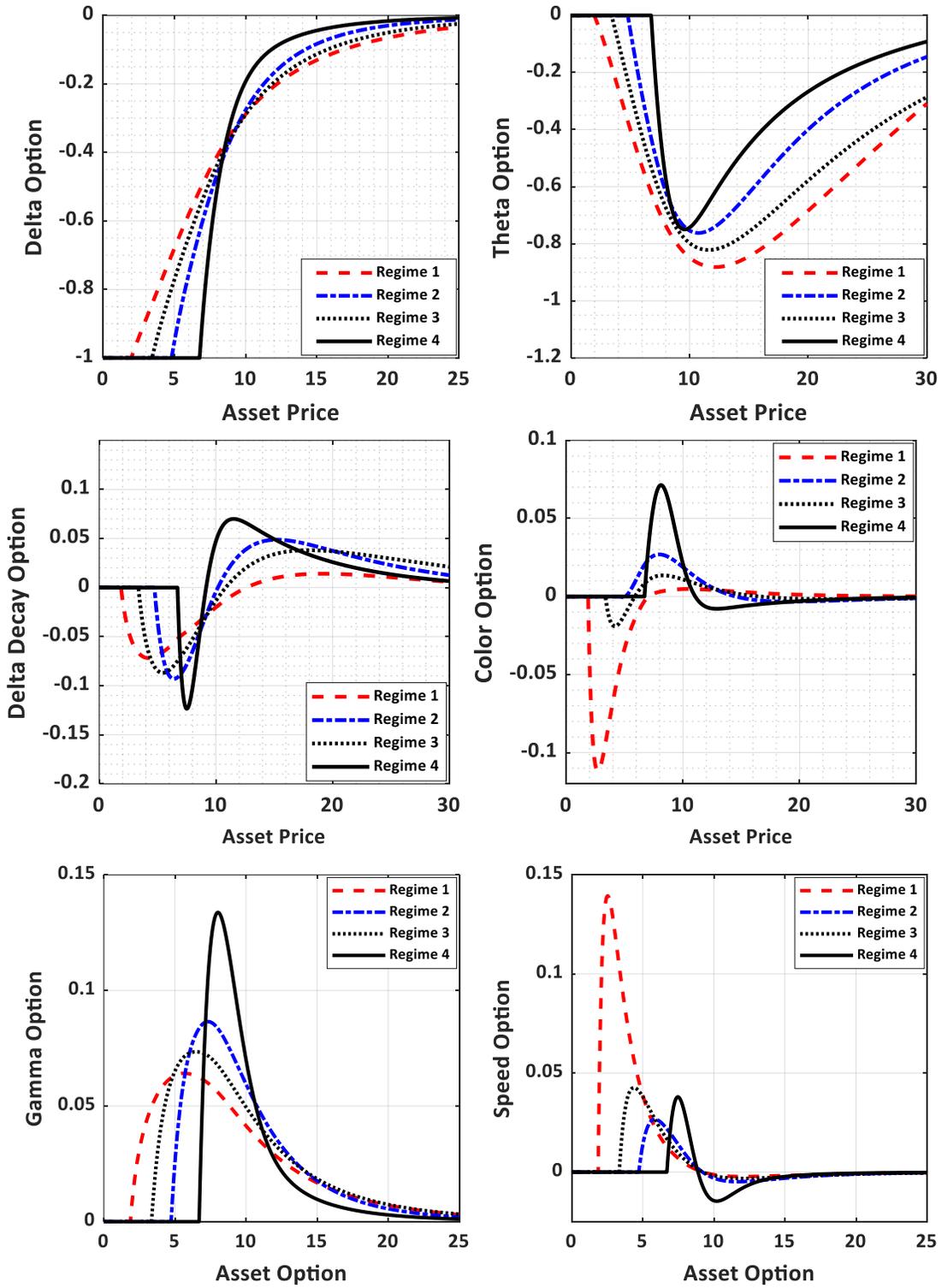

**Fig. 6.** Option Greeks for the four-regime case when $\tau = T$.

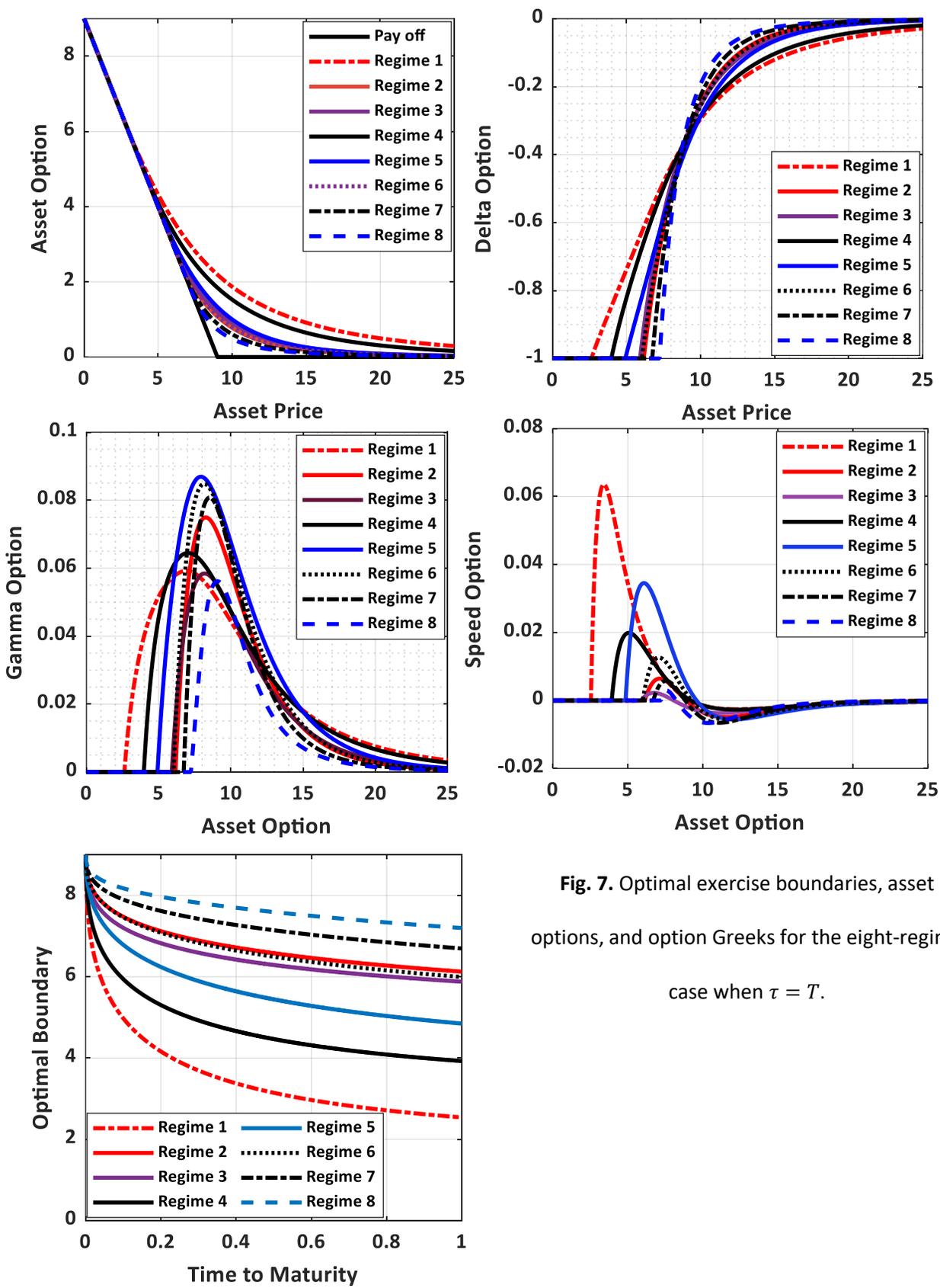

**Fig. 7.** Optimal exercise boundaries, asset options, and option Greeks for the eight-regime case when $\tau = T$.

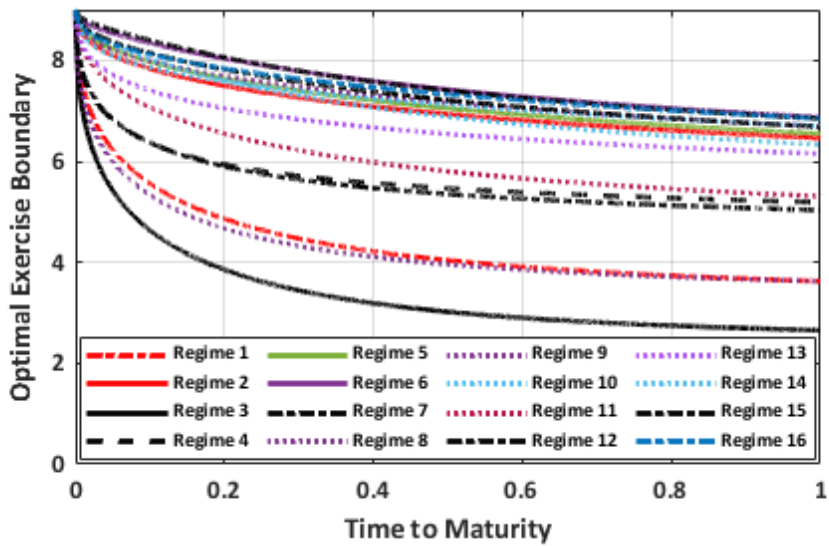

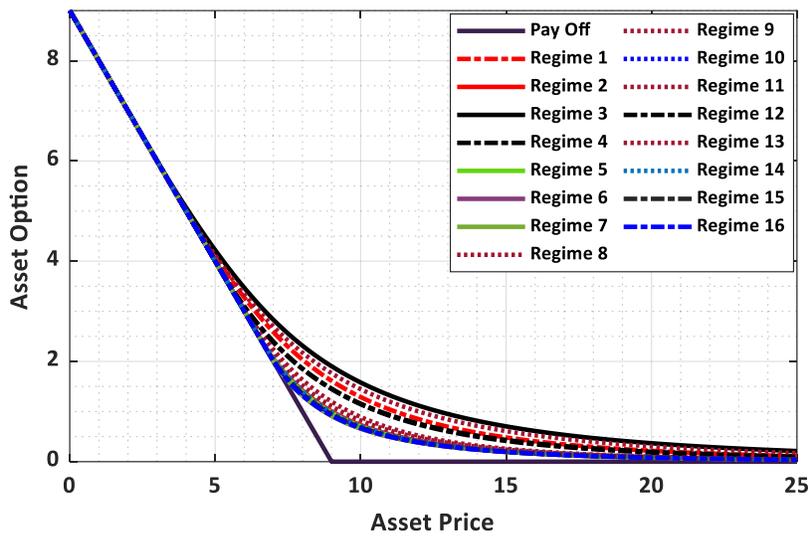

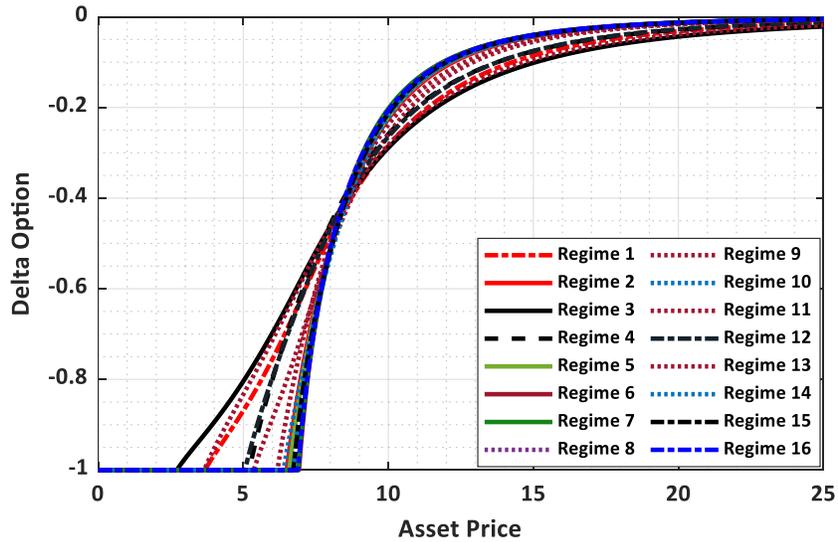

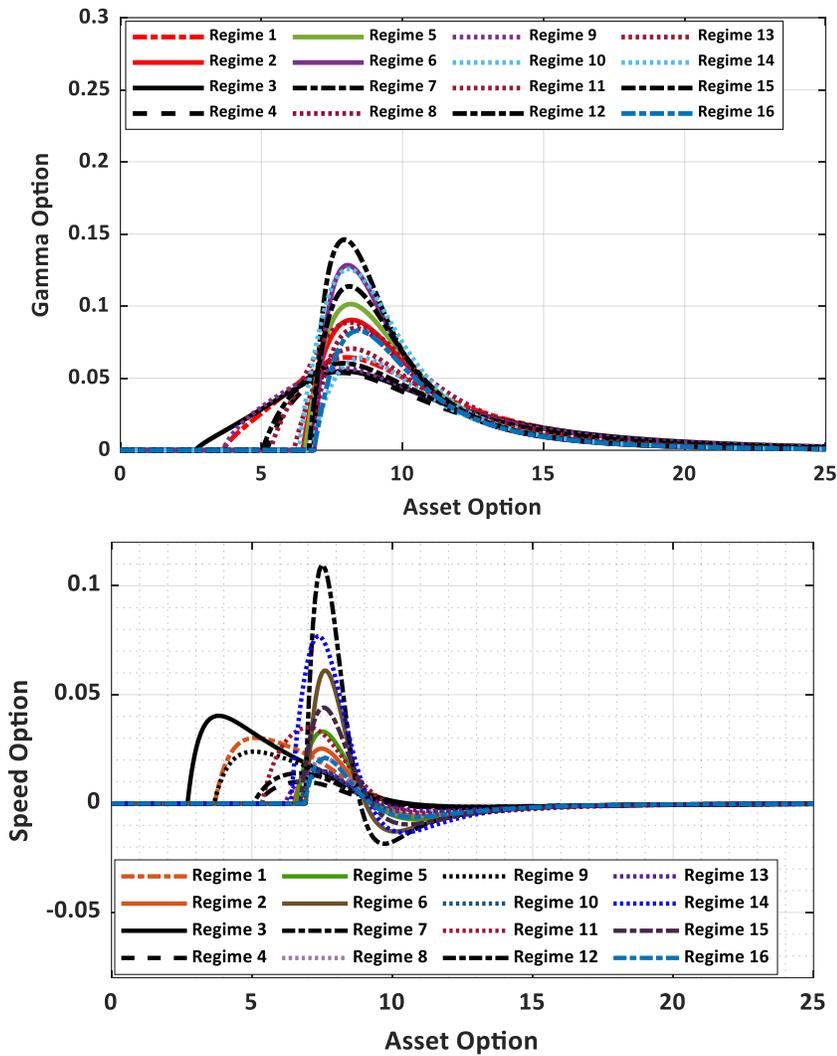

**Fig. 8.** Optimal exercise boundaries, asset options, and option Greeks for the sixteen-regime case when $\tau = T$.

**Table 11.** Comparison of American put options price for the four-regime example.

| | MTree | | | | RBF-FD | | | | FF-expl | | | |
|---|---|---|---|---|---|---|---|---|---|---|---|---|
| S | Reg 1 | Reg 2 | Reg 3 | Reg 4 | Reg 1 | Reg 2 | Reg 3 | Reg 4 | Reg 1 | Reg 2 | Reg 3 | Reg 4 |
| 7.5 | 3.1433 | 2.2319 | 2.6746 | 1.6574 | 3.1424 | 2.2320 | 2.6744 | 1.6576 | 3.1421 | 2.2313 | 2.6739 | 1.6573 |
| 9.0 | 2.5576 | 1.5834 | 2.0568 | 0.9855 | 2.5564 | 1.5835 | 2.0566 | 0.9857 | 2.5563 | 1.5827 | 2.0559 | 0.9850 |
| 10.5 | 2.1064 | 1.1417 | 1.6014 | 0.6533 | 2.1052 | 1.1415 | 1.6013 | 0.6554 | 2.1047 | 1.1406 | 1.6004 | 0.6546 |
| 12.0 | 1.7545 | 0.8377 | 1.2625 | 0.4708 | 1.7527 | 0.8377 | 1.2625 | 0.4708 | 1.7524 | 0.8368 | 1.2614 | 0.4700 |
| | ETD-CN | | | | | | | | FF-CS2 | | | |
| 7.5 | 3.1513 | 2.2384 | 2.6813 | 1.6664 | | | | | 3.1418 | 2.2319 | 2.6746 | 1.6578 |
| 9.0 | 2.5641 | 1.5884 | 2.0623 | 0.9903 | | | | | 2.5545 | 1.5835 | 2.0567 | 0.9858 |
| 10.5 | 2.1113 | 1.1451 | 1.6057 | 0.6580 | | | | | 2.1015 | 1.1414 | 1.6012 | 0.6553 |
| 12.0 | 1.7578 | 0.8377 | 1.2658 | 0.4725 | | | | | 1.7525 | 0.8374 | 1.2621 | 0.4706 |

**Table 12.** American put options price for the eight- and sixteen-regime examples using FF-CS2.

| | Eight regimes | | | | | Sixteen regimes | | | | | | |
|---|---|---|---|---|---|---|---|---|---|---|---|---|
| S | Reg 1 | Reg 2 | Reg 4 | Reg 6 | Reg 8 | Reg 1 | Reg 2 | Reg 4 | Reg 6 | Reg 8 | Reg 12 | Reg 16 |
| 3.5 | 5.5551 | 5.5000 | 5.5000 | 5.5000 | 5.5000 | 5.5000 | 5.5000 | 5.5000 | 5.5000 | 5.5000 | 5.5000 | 5.5000 |
| 4.0 | 5.1238 | 5.0000 | 5.0006 | 5.0000 | 5.0000 | 5.0074 | 5.0000 | 5.0000 | 5.0000 | 5.0000 | 5.0000 | 5.0000 |
| 4.5 | 4.7190 | 4.5000 | 4.5319 | 4.5000 | 4.5000 | 4.5385 | 4.5000 | 4.5000 | 4.5000 | 4.5000 | 4.5000 | 4.5000 |
| 6.0 | 3.6630 | 3.0000 | 3.3646 | 3.0001 | 3.0000 | 3.2833 | 3.0000 | 3.0872 | 3.0000 | 3.0000 | 3.1064 | 3.0000 |
| 7.5 | 2.8399 | 1.7960 | 2.4955 | 1.8250 | 1.5300 | 2.3075 | 1.7145 | 2.1058 | 1.6227 | 1.6624 | 2.1162 | 1.6248 |
| 8.5 | 2.4071 | 1.2861 | 2.0532 | 1.3135 | 0.9336 | 1.8260 | 1.2060 | 1.6495 | 1.1088 | 1.1533 | 1.6496 | 1.1144 |
| 9.0 | 2.2204 | 1.0918 | 1.8658 | 1.1166 | 0.7455 | 1.6287 | 1.0207 | 1.4661 | 0.9311 | 0.9730 | 1.4618 | 0.9378 |
| 9.5 | 2.0509 | 0.9290 | 1.6980 | 0.9508 | 0.6030 | 1.4560 | 0.8696 | 1.3071 | 0.7898 | 0.8272 | 1.2990 | 0.7963 |
| 10.5 | 1.7572 | 0.6782 | 1.4124 | 0.6942 | 0.4092 | 1.1718 | 0.6434 | 1.0483 | 0.5842 | 0.6112 | 1.0346 | 0.5883 |
| 12.0 | 1.4083 | 0.4332 | 1.0833 | 0.4426 | 0.2480 | 0.8614 | 0.4285 | 0.7690 | 0.3928 | 0.4079 | 0.7508 | 0.3936 |

**Table 13.** American put options price for the eight- and sixteen-regime examples using FF-CS4.

| | Eight regimes | | | | | Sixteen regimes | | | | | | |
|---|---|---|---|---|---|---|---|---|---|---|---|---|
| S | Reg 1 | Reg 2 | Reg 4 | Reg 6 | Reg 8 | Reg 1 | Reg 2 | Reg 4 | Reg 6 | Reg 8 | Reg 12 | Reg 16 |
| 3.5 | 5.5555 | 5.5000 | 5.5000 | 5.5000 | 5.5000 | 5.5000 | 5.5000 | 5.5000 | 5.5000 | 5.5000 | 5.5000 | 5.5000 |
| 4.0 | 5.1244 | 5.0000 | 5.0006 | 5.0000 | 5.0000 | 5.0075 | 5.0000 | 5.0000 | 5.0000 | 5.0000 | 5.0000 | 5.0000 |
| 4.5 | 4.7197 | 4.5000 | 4.5320 | 4.5000 | 4.5000 | 4.5387 | 4.5000 | 4.5000 | 4.5000 | 4.5000 | 4.5000 | 4.5000 |
| 6.0 | 3.6639 | 3.0000 | 3.3649 | 3.0001 | 3.0000 | 3.2836 | 3.0000 | 3.0872 | 3.0000 | 3.0000 | 3.1064 | 3.0000 |
| 7.5 | 2.8408 | 1.7962 | 2.4959 | 1.8252 | 1.5301 | 2.3078 | 1.7145 | 2.1059 | 1.6227 | 1.6611 | 2.1163 | 1.6249 |
| 8.5 | 2.4081 | 1.2864 | 2.0536 | 1.3138 | 0.9338 | 1.8264 | 1.2061 | 1.6496 | 1.1089 | 1.1555 | 1.6497 | 1.1146 |
| 9.0 | 2.2214 | 1.0921 | 1.8662 | 1.1169 | 0.7457 | 1.6290 | 1.0209 | 1.4662 | 0.9312 | 0.9767 | 1.4619 | 0.9379 |
| 9.5 | 2.0519 | 0.9293 | 1.6985 | 0.9510 | 0.6033 | 1.4563 | 0.8697 | 1.3072 | 0.7900 | 0.8319 | 1.2992 | 0.7964 |
| 10.5 | 1.7582 | 0.6785 | 1.4129 | 0.6945 | 0.4094 | 1.1721 | 0.6435 | 1.0484 | 0.5884 | 0.6114 | 1.0347 | 0.5885 |
| 12.0 | 1.4092 | 0.4334 | 1.0838 | 0.4428 | 0.2482 | 0.8617 | 0.4286 | 0.7692 | 0.3929 | 0.4081 | 0.7509 | 0.3938 |

**Table 14.** American put option Greeks for the four-regime example with FF-CS2.

| | Delta | | | Gamma | | | Speed | | |
|---|---|---|---|---|---|---|---|---|---|
| S | Reg 1 | Reg 2 | Reg 4 | Reg 1 | Reg 2 | Reg 4 | Reg 1 | Reg 2 | Reg 4 |
| 3.5 | -0.8246 | -1.0000 | -1.0000 | 0.0515 | 0.0000 | 0.0000 | 0.0949 | 0.0000 | 0.0000 |
| 6.0 | -0.5739 | -0.7442 | -1.0000 | 0.0638 | 0.0723 | 0.0000 | 0.0204 | 0.0262 | 0.0000 |
| 9.0 | -0.3401 | -0.3546 | -0.3026 | 0.0488 | 0.0727 | 0.1086 | 0.0005 | 0.0011 | -0.0027 |
| 12.0 | -0.2055 | -0.1679 | -0.0958 | 0.0289 | 0.0370 | 0.0269 | -0.0023 | -0.0048 | -0.0081 |

At the money option, volatility has a negligible impact on the delta option for all the regimes. Hence, the plot for each regime intersects at the strike price. For long put options, as we move deep in the money and out of the money, delta converges to -1 and 0, respectively. Gamma is maximum when at the money. Ignoring the sign convention, the theta of ATM is maximum. Delta decay and color options measure the rate at which delta and gamma options decay, respectively.

## 5. Conclusion

We have developed an accurate numerical method for solving American put options with regime-switching. Through the front-fixing transformation, we were able to map the optimal exercise boundary for each regime to a fixed interval. The derivative transformation enables us to employ the higher-order compact finite difference method coupled with the Hermite interpolation for solving the system of the asset, delta, gamma, and speed options while capturing the optimal exercise boundary and theta, delta decay and color options. Moreover, our method has a substantial advantage because it simultaneously and accurately calculates asset, delta, gamma, speed, theta, delta decay, and color options, as well as optimal exercise boundary during iteration. Greek parameters are difficult to estimate correctly as can be seen from previous works of literature. However, by formulating a set of systems of PDEs that consist of the asset option and its derivatives for each regime, we were able to estimate those parameters with higher-order accuracy. Our numerical discretization also presents a system where the coefficient matrix is tridiagonal and positive definite with constant-elements, which enables us to implement both Gauss-Seidel and Newton iterations (with Thomas algorithm) with simple computation. The present scheme has been tested in two-, four-, eight-, and sixteen-regime problems and with cubic and quintic Hermite interpolations. The results show that the method provides an accurate solution and is fast in computation as compared with the existing methods. Future research will include applying this method to non-constant volatility and/or interest rate cases.

## References

1. Adam, Y. (1975). A Hermitian finite difference method for the solution of parabolic equation. Computers and Mathematics with Applications, 1, 393-406.
2. Adam, Y. (1976). Highly accurate compact implicit methods and boundary condition. Journal of Computational Physics, 24, 10-22.
3. Babbin, J., Forsyth, P. A., and Labahn, G. (2014). A comparison of iterated optimal stopping and local policy iteration for American options under regime switching. Journal of Scientific Computing, 58, 409-430.
4. Bidégaray-Fesquet, B. (2006). Von-Neumann stability analysis of FD-TD methods in complex media.
5. Bingham, N. H. (2006). Financial modeling with jump processes. Journal of the American Statistical Association, 101 (475), 1315-1316 (https://doi:10.1198/jasa.2006.s130).

6. Blackwell, B. F., and Hogan, R. E. (1994). One-dimensional ablation using Landau transformation and finite control volume procedure. Journal of Thermophysics and Heat Transfer, 8 (2), 282-287 (https://doi:10.2514/3.535).

7. Boyarchenko, S. I., and Levendorskii, S. Z. (2008). Pricing American options in regime-switching models: FFT Realization. SSRN Electronic Journal (https://doi:10.2139/ssrn.1127562).

8. Bravo, E. G., and Mcgraw, T. (2015). Visualizing Aldo Giorgini's ideal's flow. Springer International Publishing. Doi:10.1007/978-3-319-27863-6_72.

9. Buffington, J., and Elliott, R. J. (2002). American options with regime-switching. International Journal of Theoretical and Applied Finance, 5 (05), 497-514 (https://doi:10.1142/s0219024902001523).

10. Burden, R. L., Faires, D. J., and Burden, A. M. (2016). Numerical Analysis. Cengage Learning, Boston, MA.

11. Cao, H. H., Liu, L. B., Zhang, Y., and Fu, S. M. (2011). A fourth-order method of the convection-diffusion equation with Neumann boundary conditions. Applied Mathematics and Computation, 217, 9133-9141.

12. Chapra, S. C. (2012). Applied Numerical Methods with MATLAB for Engineers and Scientist. MC Graw-Hill, New York.

13. Chen, Y., Xiao, A., and Wang, W. (2019). An IMEX-BDF2 compact scheme for pricing options under regime-switching jump-diffusion models. Mathematical Methods in the Applied Sciences, (https://doi:10.1002/mma.5539).

14. Chiarella, C., Nikitopoulos Sklibosios, C., Schlogl, E., and Yang, H. (2016). Pricing American options under regime switching using method of lines. SSRN Electronic Journal, (https://doi:10.2139/ssrn.2731087).

15. Chockalingam, A., and Muthuraman, K. (2011). American options under stochastic volatility. Operations Research, 59(4), 793-809 (https://doi:10.1287/opre.1110.0945).

16. Company, R., Egorova, V.N., and Jódar, L. (2016). A positive, stable and consistent front-fixing numerical scheme for American Options. Springer International Publishing AG 2016 (https://doi 10.1007/978-3-319-23413-7_10).

17. Company, R., Egorova, V., Jódar, L., and Vázquez, C. (2016). Computing American option price under regime switching with rationality parameter. Computers & Mathematics with Applications, 72 (3), 741-754 (https://doi:10.1016/j.camwa.2016.05.026).
37

18. Company, R., Egorova, V. N., and Jódar, L. (2016). Constructing positive reliable numerical solution for American call options: A new front-fixing approach. Journal of Computational and Applied Mathematics, 291, 422-431 (https://doi:10.1016/j.cam.2014.09.013).

19. Company, R., Egorova, V., Jódar, L., and Vázquez, C. (2016). Finite difference methods for pricing American put option with rationality parameter: Numerical analysis and computing. Journal of Computational and Applied Mathematics, 304, 1-17 (https://doi:10.1016/j.cam.2016.03.001).

20. Cont., R., and Tankov, P. (2004). Financial Modelling with Jump Diffusion. Chapman & Hall, London.

21. Crank, J. (1984). Free and moving boundary problem, Oxford University Press, London.

22. Dremkova, E., and Ehrhardt, M. (2011). A high-order compact method for non-linear Black-Scholes option pricing equation of American options. International Journal of Computational Mathematics, 88, 2782-2797.

23. Düring, B., and Fournié, M. (2012). High-order compact finite difference scheme for option pricing in stochastic volatility models. Journal of Computational and Applied Mathematics, 236 (17), 4462-4473 (https://doi:10.1016/j.cam.2012.04.017).

24. Egorova, V. N., Company, R., and Jódar, L. (2016). A new efficient numerical method for solving American option under regime switching model. Computers and Mathematics with Applications, 71 (1), 224–237 (https://doi:10.1016/j.camwa.2015.11.019).

25. Elliott, R. J., Siu, T. K., Chan, L., and Lau, J. W. (2007). Pricing options under a generalized markov-modulated jump-diffusion model. Stochastic Analysis and Applications, 25 (4), 821–843.doi:10.1080/07362990701420118 .

26. Gao, G. and Sun, Z. (2013). Compact difference schemes for heat equation with Neumann boundary condition (II), Numerical Methods for Partial Differential Equations, 29 (5), 1459-1486.

27. Garnier, J., and Sølna, K. (2017). Correction to Black--Scholes formula due to fractional stochastic volatility. SIAM Journal on Financial Mathematics, 8 (1), 560–588 (https://doi:10.1137/ 15m1036749 Stochastic Volatility Correction to Black-Scholes).

28. Guoqing, Y., and Hanson, F. B. (2006). Option pricing for a stochastic-volatility jump-diffusion model with log-uniform jump-amplitudes. American Control Conference, (https://doi:10.1109/ acc.2006.1657175).

29. Hamilton, J. D. (1989). A new approach to the economic analysis of nonstationary time series and the business cycle. Econometrica, 57 (2), 357-384 (https://doi:10.2307/1912559).
38
18. Company, R., Egorova, V. N., and Jódar, L. (2016). Constructing positive reliable numerical solution for American call options: A new front-fixing approach. Journal of Computational and Applied Mathematics, 291, 422-431 (https://doi:10.1016/j.cam.2014.09.013).

19. Company, R., Egorova, V., Jódar, L., and Vázquez, C. (2016). Finite difference methods for pricing American put option with rationality parameter: Numerical analysis and computing. Journal of Computational and Applied Mathematics, 304, 1-17 (https://doi:10.1016/j.cam.2016.03.001).

20. Cont., R., and Tankov, P. (2004). Financial Modelling with Jump Diffusion. Chapman & Hall, London.

21. Crank, J. (1984). Free and moving boundary problem, Oxford University Press, London.

22. Dremkova, E., and Ehrhardt, M. (2011). A high-order compact method for non-linear Black-Scholes option pricing equation of American options. International Journal of Computational Mathematics, 88, 2782-2797.

23. Düring, B., and Fournié, M. (2012). High-order compact finite difference scheme for option pricing in stochastic volatility models. Journal of Computational and Applied Mathematics, 236 (17), 4462-4473 (https://doi:10.1016/j.cam.2012.04.017).

24. Egorova, V. N., Company, R., and Jódar, L. (2016). A new efficient numerical method for solving American option under regime switching model. Computers and Mathematics with Applications, 71 (1), 224–237 (https://doi:10.1016/j.camwa.2015.11.019).

25. Elliott, R. J., Siu, T. K., Chan, L., and Lau, J. W. (2007). Pricing options under a generalized markov-modulated jump-diffusion model. Stochastic Analysis and Applications, 25 (4), 821–843.doi:10.1080/07362990701420118 .

26. Gao, G. and Sun, Z. (2013). Compact difference schemes for heat equation with Neumann boundary condition (II), Numerical Methods for Partial Differential Equations, 29 (5), 1459-1486.

27. Garnier, J., and Sølna, K. (2017). Correction to Black--Scholes formula due to fractional stochastic volatility. SIAM Journal on Financial Mathematics, 8 (1), 560–588 (https://doi:10.1137/ 15m1036749 Stochastic Volatility Correction to Black-Scholes).

28. Guoqing, Y., and Hanson, F. B. (2006). Option pricing for a stochastic-volatility jump-diffusion model with log-uniform jump-amplitudes. American Control Conference, (https://doi:10.1109/ acc.2006.1657175).

29. Hamilton, J. D. (1989). A new approach to the economic analysis of nonstationary time series and the business cycle. Econometrica, 57 (2), 357-384 (https://doi:10.2307/1912559).
38

30. Han, Y., and Kim, G. (2016). Efficient lattice method for valuing of options with barrier in a regime-switching model. Discrete Dynamics in Nature and Society, 2016, 1-14 (https://doi:10.1155/2016/2474305).

31. Hirsch, C. (2001). Numerical Computation of Internal and External Flows: Fundamental of Numerical Discretization. Wiley, New York.

32. Hirsh, R. S. (1975). Higher order accurate difference solutions of fluid mechanics by a compact differencing technique. Journal of Computational Physics, 19, 90-109.

33. Huang, Y., Forsyth, P. A., and Labahn, G. (2011). Methods for pricing american options under regime switching. SIAM Journal on Scientific Computing, 33 (5), 2144-2168 (https:// doi:10.1137/110820920).

34. Hull, J., and White, A. (1987). The pricing of options on assets with stochastic volatilities. The Journal of Finance, 42 (2), 281-300.

35. Ikonen, S., and Toivanen, J. (2007). Efficient numerical methods for pricing American options under stochastic volatility. Numerical Methods for Partial Differential Equations, 24 (1), 104-126 (https://doi:10.1002/num.20239).

36. Kangro, R., and Nicolaides, R. (2000). Far field boundary conditions for Black--Scholes equations. SIAM Journal on Numerical Analysis, 38 (4), 1357-1368 (https://doi:10.1137/ s0036142999355921).

37. Khaliq, A. Q. M., and Liu, R. H. (2009). New numerical scheme for pricing American option with regime-switching. International Journal of Theoretical and Applied Finance, 12 (03), 319-340 (https://doi:10.1142/s0219024909005245).

38. Khaliq A. Q. M., Kleefeld, B., and Liu, R. H. (2013). Solving complex PDE systems for pricing American options with regime-switching by efficient exponential time differencing schemes. Numerical Methods Partial Differential Equation, 29 (1), 320-336.

39. Kou, S. G. (2002). A jump-diffusion model for option pricing. Management Science, 48 (8), 1086-1101 (https://doi:10.1287/mnsc.48.8.1086.166).

40. Kwok Y. K. (2008). Mathematical Models of Financial Derivatives. Springer, Berlin.

41. Landau, H. G. (1950). Heat conduction in a melting solid. Quarterly of Applied Mathematics, 8 (1), 81-94 (https://doi:10.1090/qam/33441).

42. Li, H., Mollapourasi, R., and Haghi, M. (2018). A local radial basis function method for pricing options under the regime switching model. Journal of Scientific Computing, 79, 517-541.

43. Liao, W., Zhu, J., and Khaliq, A. Q. M. (2002). An efficient high-order algorithm for solving systems of reaction-diffusion equations. Numerical Methods for Partial Differential Equations, 18 (3), 340-354.

44. Liao, W., and Khaliq, A. Q. M. (2009). High-order compact scheme for solving nonlinear Black-Scholes equation with transaction cost. International Journal of Computer Mathematics, 86 (6), 1009-1023 (https://doi:10.1080/00207160802609829).

45. Liu, R. H. (2010). Regime-switching recombining tree for option pricing. International Journal of Theoretical and Applied Finance, 13, 479-499.

46. Liu, R. H., Zhang, Q., and Yin, G. (2006). Option pricing in a regime-switching model using the fast Fourier transform. Journal of Applied Mathematics and Stochastic Analysis, 2006, 1-22 (https://doi:10.1155/jamsa/2006/18109).

47. Mamon, R. S., and Rodrigo, M. R. (2005). Explicit solutions to European options in a regime-switching economy. Operations Research Letters, 33 (6), 581-586 (https://doi:10.1016/ j.orl.2004.12.003).

48. Meyer, G. H., and van der Hoek, J. (1997). The evaluation of American options with the method of lines. Advances in Futures and Options Research, 9 (9), 265-285.

49. Mitchell, S. L., and Vynnycky, M. (2009). Finite-difference methods with increased accuracy and correct initialization for one-dimensional Stefan problems. Applied Mathematics and Computation, 215 (4), 1609-1621 (https://doi:10.1016/j.amc.2009.07.054).

50. Mitchell, S. L., and Vynnycky, M. (2012). An accurate finite-difference method for ablation-type Stefan problems. Journal of Computational and Applied Mathematics, 236 (17), 4181-4192 (https://doi:10.1016/j.cam.2012.05.011).

51. Morton, W.K., and Mayer, D. (2005). Numerical Solutions of Partial Differential Equations. Cambridge University Press, London.

52. Nielsen, B. F., Skavhaug O., and Tveito A. (2002). A penalty and front-fixing methods for the numerical solution of American option problems. Journal of Computational Finance, 5 (4), 69-97.

53. Norris, J. R. (1998). Markov Chain. Cambridge University Press, London.

54. Shang, Q., and Bryne, B., (2019). An efficient lattice search algorithm for the optimal exercise boundary in American options. SSRN Electronic Journal.

55. Tauryawati, M. L., Imron, C., and Putri, E. R. (2018). Finite volume method for pricing European call option with regime-switching volatility. Journal of Physics: Conference Series, 974, (https://doi:10.1088/1742-6596/974/1/012024).

56. Toivanen, J. (2010). Finite difference methods for early exercise options. Encyclopedia of Quantitative Finance, (https://doi:10.1002/9780470061602.eqf12002).

57. Ugur, O. (2009). Introduction to Computational Finance. Imperial College Press, London. (https://doi.org/10.1142/p556).

58. Wu, L., and Kwok, Y.K. (1997). A front-fixing method for the valuation of American option, Journal of Financial Engineering, 6 (2), 83-97.

59. Yan, Y., Dai, W., Wu, L., and Zhai, S. (2019). Accurate gradient preserved method for solving heat conduction equations in double layers. Applied Mathematics and Computation, 354, 58-85.

60. Zhang, K., Teo, K. L., and Swartz, M. (2013). A Robust numerical scheme for pricing American options under regime switching based on penalty method. Computational Economics, 43 (4), 463-483 (https://doi:10.1007/s10614-013-9361-3).

61. Zhao, J., Davidson, M., and Corless, R. M. (2007). Compact finite difference method for American option pricing. Journal of Computational and Applied Mathematics, 206, 306-321.

62. Zhylyevskyy, O. (2009). A fast Fourier transform technique for pricing American options under stochastic volatility. Review of Derivatives Research, 13 (1), 1-24 (https://doi:10.1007/s11147-009-9041-6).